 \documentclass[final,5p,times,twocolumn,authoryear]{elsarticle}

\usepackage{amssymb}
\usepackage{amsmath}

\usepackage{booktabs,caption}
\usepackage{threeparttable}
\usepackage{multirow}
\usepackage{array, bigdelim, makecell} 
\usepackage{siunitx}

\journal{Icarus}

\graphicspath{ {./figures/} }

\begin{document}

\begin{frontmatter}



\title{Energy deposition in Saturn's equatorial upper atmosphere}


\author[imperial]{J. M. Chadney}
\author[lpl]{T. T. Koskinen}
\author[imperial]{X. Hu}
\author[imperial]{M. Galand}
\author[gsma]{P. Lavvas}
\author[imperial]{Y. C. Unruh}
\author[hopkins]{J. Serigano}
\author[hopkins]{S. M. H\"{o}rst}
\author[lpl]{R. V. Yelle}

\address[imperial]{Department of Physics, Imperial College London, Prince Consort Road, London SW7 2AZ, UK}
\address[lpl]{Lunar and Planetary Laboratory, University of Arizona, 1629 E. University Blvd., Tucson, AZ 85721, USA}
\address[gsma]{GSMA, Universit\'{e} de Reims Champagne Ardenne, CNRS UMR 7331, 51687 Reims, France}
\address[hopkins]{Department of Earth and Planetary Sciences, Johns Hopkins University, Baltimore, MD, USA}

\begin{abstract}
We construct Saturn equatorial neutral temperature and density profiles of H, H$_2$, He, and CH$_4$, between 10$^{-12}$ and 1~bar using measurements from Cassini's Ion Neutral Mass Spectrometer (INMS) taken during the spacecraft's final plunge into Saturn's atmosphere on 15 September 2017, combined with previous deeper atmospheric measurements from the Cassini Composite InfraRed Spectrometer (CIRS) and from the UltraViolet Imaging Spectrograph (UVIS). These neutral profiles are fed into an energy deposition model employing soft X-ray and Extreme UltraViolet (EUV) solar fluxes at a range of spectral resolutions ($\Delta\lambda=4\times10^{-3}$~nm to 1~nm) assembled from TIMED/SEE, from SOHO/SUMER, and from the Whole Heliosphere Interval (WHI) quiet Sun campaign. Our energy deposition model calculates ion production rate profiles through photo-ionisation and electron-impact ionisation processes, as well as rates of photo-dissociation of CH$_4$. The ion reaction rate profiles we determine are important to obtain accurate ion density profiles, meanwhile methane photo-dissociation is key to initiate complex organic chemical processes. We assess the importance of spectral resolution in the energy deposition model by using a high-resolution H$_2$ photo-absorption cross section, which has the effect of producing additional ionisation peaks near 800~km altitude. We find that these peaks are still formed when using low-resolution ($\Delta\lambda=1$~nm) or mid-resolution ($\Delta\lambda=0.1$~nm) solar spectra, as long as high-resolution cross sections are included in the model.

\end{abstract}

\end{frontmatter}

\section{Introduction}

Between April and September 2017, after spending 13 years exploring the Kronian system, the Cassini spacecraft performed 23 proximal orbits passing within Saturn's D ring, allowing us to directly probe the upper atmosphere of a gas giant planet. This ``Grand Finale'' phase of the mission culminated in a ``Final Plunge'' into Saturn's atmosphere on 15 September 2017 during which in-situ measurements were taken until the signal from the spacecraft was lost at an altitude of approximately 1360~km above the 1~bar level. The proximal orbits probed Saturn's equatorial region, with the final plunge spanning latitudes from about 13$^\circ$N to 9$^\circ$N as Cassini descended from 2500~km to below 1400~km altitude \citep{Waite2018}.

A number of studies have detailed the influence of the rings on Saturn's upper atmosphere revealed by the Grand Finale observations. \citet{Wahlund2018} used measurements from the Radio and Plasma Wave Science instrument (RPWS) to derive electron densities during the proximal orbits. Large variations in electron density of up to two orders of magnitude were measured, induced by shadows from the rings reducing ionisation. Despite high variability between proximal orbits, \citet{Hadid2019} were able to measure a consistent difference in electron density derived from the RPWS Langmuir probe between the northern and southern hemispheres, explained by the influence of the rings. Variable response in the ionosphere linked to the B ring was interpreted as inter-hemispheric transport from the sunlit to the shadowed ionosphere in the light of INMS ion observations.

The Ion Neutral Mass Spectrometer (INMS) measured neutral atmospheric composition down to the pressure level of about 1~nbar during the final plunge. \citet{Yelle2018} analysed the low-mass end of the INMS neutral spectrum to derive the number densities of H$_2$, He, CH$_4$, as well as the neutral temperature profile. They found that H$_2$ and He were in diffusive equilibrium but that there were high volume mixing ratios of CH$_4$ above the homopause, the latter with a mole fraction on the order of a few times 10$^{-4}$. The slope of the CH$_4$ mixing ratio was consistent with inflow from above the atmosphere, possibly from the rings. Atmospheric constituents from the high-mass region of the INMS spectrum measured during the final plunge have been studied by \citet{Waite2018}. Evidence of inflows of other molecules, such as CO and N$_2$, were found in the data, with possible source regions in the D ring identified. Before Cassini's final plunge, the only inflow from the rings that was known was H$_2$O, inferred from electron density measurements \citep[e.g.,][]{Jurac2005,Prange2006,Moore2006a}.

There is evidence that inflow of heavy species from the rings has an impact on ion composition. \citet{Cravens2019} found that INMS recorded lower than expected quantities of the ions H$^+$ and H$_3^+$ during the proximal orbits. These light ions may have been destroyed in reactions with inflowing neutral molecules. Ionospheric modelling by \citet{Moore2018} confirms the destruction of light ions by heavy molecular species, with the resulting ionosphere containing large volume mixing ratios of molecular ions, such as H$_3$O$^+$, HCO$^+$. However, their model was not able to reproduce the number density of H$_3^+$; the calculated density was found to be too large in comparison to INMS data, indicating missing loss processes for this species in the ionosphere.

In this study, we combine the INMS measurements during the final plunge with previous limb scans from the Cassini Composite InfraRed Spectrometer (CIRS) and stellar occultation measurements by the UltraViolet Imaging Spectrograph (UVIS) to obtain the most realistic profiles in altitude of the temperature and major neutral densities (H$_2$, H, He, CH$_4$) of the equatorial upper atmosphere down to the 1~bar pressure level (set by convention to z=0). We use this neutral upper atmosphere in an energy deposition model to predict ionisation rates under solar illumination, including photo-ionisation and electron-impact ionisation. We also determine photo-dissociation rate profiles of methane, which is key to initiate the chemical reactions leading to the formation of more complex hydrocarbons. We make use of high spectral resolution solar fluxes combined with a high resolution H$_2$ photo-absorption cross-section in order to test the importance of spectral resolution in Saturn ionospheric models. Some previous studies have included high resolution photo-absorption cross sections of H$_2$ at Saturn \citep{Kim2014} and Jupiter \citep{Kim1994}, and of N$_2$ at Titan \citep{Lavvas2011}; beyond the ionisation threshold wavelength (80.4~nm for H$_2$, 79.7~nm for N$_2$), the cross section is highly structured due to excitation processes. These studies found that in the higher resolution models photons penetrate deeper into the upper atmosphere, resulting in a significant increase in H and CH$_4$ ionisation. These species can be ionised over the highly structured photoabsorption of H$_2$ and N$_2$ as their ionisation threshold wavelength is 91.2~nm and 98.8~nm, respectively.

The neutral upper atmospheric composition and temperature profiles along with ion production rates that we calculate in this paper will be important to determine accurate ion densities. Taking into account the H$_2$O influx \citep{Connerney1986,ODonoghue2013} will be necessary at this stage since water plays a critical role in  ion-neutral chemistry. In addition, the photo-dissociation of methane is key to initiate the chemical reactions leading to the formation of more complex hydrocarbons such as benzene \citep{Koskinen2016}.

Our paper is laid out as follows. In Sect.~\ref{sec:model_inputs}, we describe all of the inputs of our energy deposition model. Section~\ref{sec:ionospheric_model} describes the energy deposition model itself, and in Sect.~\ref{sec:results} we present and interpret our results. In Sect.~\ref{sec:conc}, we highlight the main findings and discuss their implications.

\section{Key model inputs}\label{sec:model_inputs}
There are a number of key inputs to our energy deposition model that we must assemble: the neutral temperature and density profiles (Sect.~\ref{sec:neutral_atm}) and the solar flux (Sect.~\ref{sec:solar_flux}), which we seek to acquire at a spectral resolution high enough to capture the structured region of the H$_2$ photo-absorption cross section (Sect.~\ref{sec:sigma_H2}). Other model inputs (reaction rates and remaining cross sections) are presented in Sect.~\ref{sec:ionospheric_model}.

\subsection{Reconstructed neutral atmosphere during Cassini plunge}\label{sec:neutral_atm}
We reconstruct the neutral atmosphere (temperature and densities of H$_2$, H, He, CH$_4$) during the final plunge of the Cassini spacecraft. To this effect, we use the deepest in-situ measurements taken during the final plunge by INMS. The INMS final plunge measurements provide densities of H$_2$, He, and CH$_4$, at pressures up to 1~nbar. To reconstruct the neutral atmosphere down to the 1~bar pressure level, we rely on previously combined CIRS limb scans and UVIS stellar occultation observations \citep{Koskinen2018}.

\subsubsection*{Neutral temperature profile}
We combine temperature measurements from CIRS and UVIS taken close to Saturn's equator with INMS temperatures taken during the final plunge. To correct for differences in latitude between the observations, we take a similar approach to \citet{Yelle2018} and use the effective potential (i.e., the sum of the gravitational and centrifugal potentials) as a vertical coordinate. Indeed, given Saturn's oblate shape and rapid rotation, the change in atmospheric parameters is not purely radial, but contains a latitudinal component. If we express the parameters as a function of the effective potential, we can assemble a temperature profile that is independent of latitude. We define the effective potential $\Phi$ as the sum of the gravitational potential and the centrifugal potential \citep{Helled2015}:
\begin{equation}\label{eqn:potential}
\Phi = \frac{GM}{r}\left(1-\sum_{n=1}^{4}J_{2n}P_{2n}(\sin\theta)\left(\frac{r_{\text{eq}}}{r}\right)^{2n}\right) + \frac{1}{2}\left(r\Omega\cos\theta\right)^2,
\end{equation}
where $G$ is the gravitational constant, $M$ is the mass of Saturn, $r$ is the radial distance, $\theta$ is the latitude, $\Omega$ is the angular velocity, $J_{2n}$ are the expansion coefficients and $P_{2n}$ are the Legendre polynomials. Note that this representation of the gravity field is not consistent with the Grand Finale data presented in \citet{Iess2019} and \citet{Militzer2019} but it will not have a major effect on our conclusions regarding energy deposition. Indeed, as shown in \citet{Koskinen2021}, improved estimates on the interior rotation rate and the detection of differential rotation from the Cassini Grand Finale observations result in a maximum difference of 1.1\% to the graviational accelation, compared to an acceleration based on the potential used in this paper.

\begin{figure*}
\centering
\includegraphics[width=.9\textwidth]{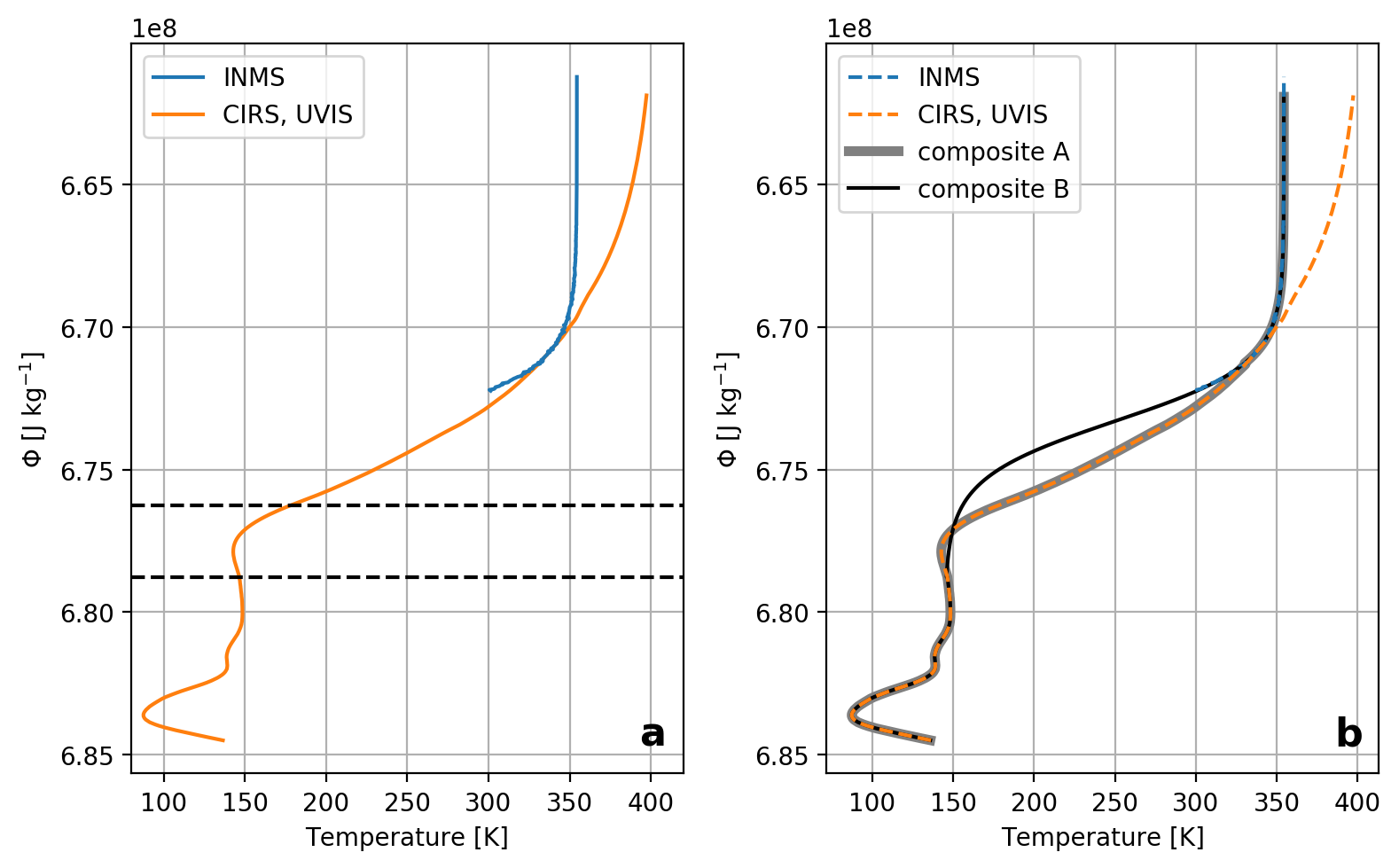}
\caption{Temperature profile measured by INMS during the final plunge (in blue), and profile assembled by \citet{Koskinen2018} (in orange) as a best fit to CIRS limb scans at $\Phi>6.788\times 10^8$~J~kg$^{-1}$ ($p>3~\mu$bar, lower dashed black line in panel a) and UVIS occultation ST14M10D03S7 at $\Phi<6.763\times 10^8$~J~kg$^{-1}$ ($p<0.04~\mu$bar, upper dashed black line in panel a). The constructed temperature profiles connecting these datasets are shown in panel b, labelled composite A (in a thick grey line), and composite B (in a thin black line).}
\label{fig:Tsource}
\end{figure*}

The temperature profile from INMS during the final plunge as a function of the potential is shown in blue in Fig.~\ref{fig:Tsource}a \citep{Yelle2018}. To extend this profile down to the 1~bar pressure level (corresponding to $\Phi=6.845\times 10^8$~J~kg$^{-1}$), we make use of the temperature profile derived by \citet{Koskinen2018}, shown in orange in Fig.~\ref{fig:Tsource}a. The latter profile is determined as a best fit to UVIS and CIRS observations in the equatorial region; specifically, UVIS occultation ST14M10D03S7, measured at a planetographic latitude of 7.4$^{\circ}$S, and CIRS limb scans LIMBINTC001 and LIMBINT001, measured at planetographic latitudes spanning 10$^{\circ}$S -- 5$^{\circ}$S, and 15$^{\circ}$S -- 2$^{\circ}$N, respectively.

The final plunge exospheric temperature measured by INMS of 354~K is consistent with the range of low latitude exospheric temperatures derived by \citet{Koskinen2018} from Cassini UVIS occultations, as well as the range of measurements and model results assembled by \citet{MullerWodarg2019}. Since the INMS profile was an in-situ observation during the final plunge, we make use of the exospheric temperature from this measurement in all of our model runs rather than that derived from the equatorial UVIS occultation profile in \citet{Koskinen2018}.

We construct two different temperature profiles to connect the INMS final plunge exospheric temperature to CIRS and UVIS equatorial observations: we label these reconstructed profiles composite A and B (plotted in Fig.~\ref{fig:Tsource}b). Composite A is made up of the CIRS and UVIS temperatures from \citet{Koskinen2018} at potentials higher than $6.713\times 10^8$~J~kg$^{-1}$ (where the two profiles intersect), and the INMS final plunge temperatures at potentials lower than this value. Since this involves discarding the lowest altitude points of the INMS observations (with potentials between $6.713\times 10^8$ and $6.722\times 10^8$~J~kg$^{-1}$), we also construct composite B which connects the INMS temperature profile to the CIRS-derived temperature region at $\Phi>6.788\times 10^8$~J~kg$^{-1}$ using a Bates profile. The parameters of the Bates temperature profile (Equation~1 from \citet{Yelle1996}) used here are chosen to best fit the INMS final plunge temperatures and to connect to the uppermost temperatures from the CIRS scans. Composite profile B ignores temperature constraints from UVIS measurements between $\Phi=6.788$ -- $6.713\times 10^8$~J~kg$^{-1}$.

\subsubsection*{Diffusion model for neutral species}
\begin{figure}
\centering
\includegraphics[width=.45\textwidth]{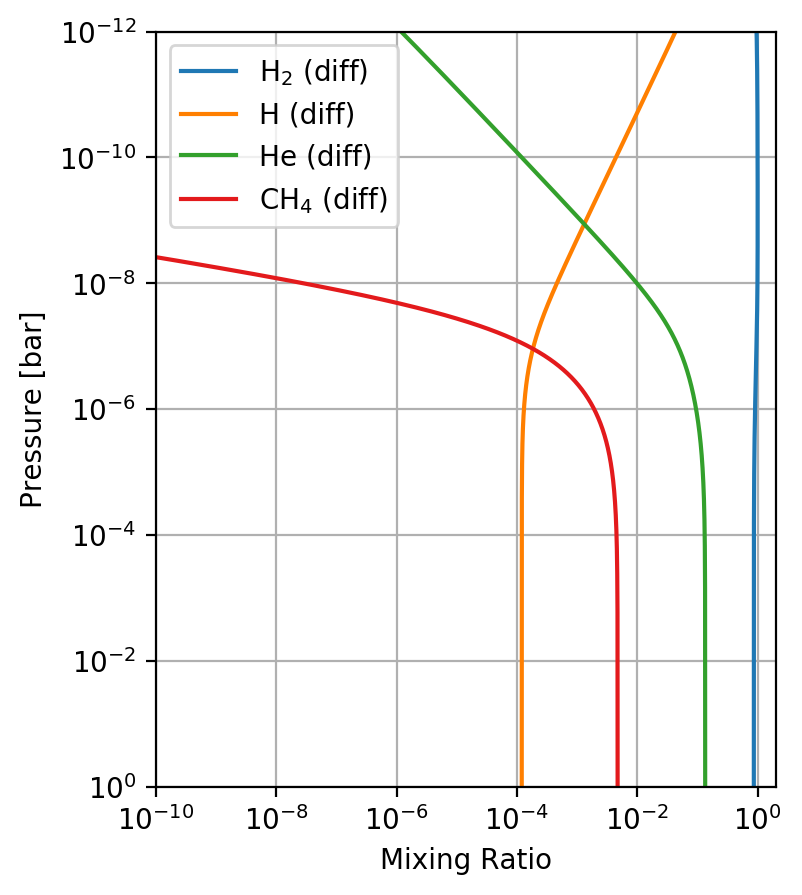}
\caption{Mixing ratios of the neutral species as a function of pressure from the diffusion model using composite temperature A.}
\label{fig:mixing_ratio_vs_p_diffusion}
\end{figure}

To reconstruct the equatorial neutral density profiles at the time of the final plunge, we use the diffusion model described in \citet{Koskinen2018}. We include the dominant neutral species in the model: H$_2$, H, He, and CH$_4$. The details of this calculation can be found in \ref{sec:appendix}. The resulting mixing ratios from the diffusion model using temperature profile A are shown as a function of pressure in Fig.~\ref{fig:mixing_ratio_vs_p_diffusion}. 

To initialise the calculation, we need the volume mixing ratios at the 1 bar pressure level $x_{s0} = x_s(z=0)$ for each neutral species $s$. For CH$_4$, we use a value of $x_{\text{CH}_4,0} = 4.7\times 10^{-3}$ from \citet{Fletcher2009}, and for H, we assume $x_{\text{H},0} = 1.2\times 10^{-4}$, resulting in a volume mixing ratio of H that is less than 0.05 in the thermosphere, in agreement with \citet{Koskinen2013}. The lower boundary volume mixing ratio of He is chosen so that the He profile in the thermosphere matches the He volume mixing ratio from the INMS final plunge measurement: we obtain $x_{\text{He},0} = 0.134$ when using composite temperature profile A, and $x_{\text{He},0} = 0.120$ with composite temperature B.

\subsubsection*{INMS final plunge volume mixing ratios}
INMS records the number densities $n_s^{\text{INMS}}$ of the neutral species $s$, namely H$_2$, He, and CH$_4$. To obtain consistent volume mixing ratios of the INMS final plunge measurements, the volume mixing ratio $x_{\text{H}}^{\text{diff}}$ of H from the diffusion model is taken into account according to
\begin{equation}\label{eqn:ntotINMS}
n_{\text{tot}}^{\text{INMS}} = \frac{1}{1-x_{\text{H}}^{\text{diff}}} \sum_s n_s^{\text{INMS}}.
\end{equation}
$n_{\text{tot}}^{\text{INMS}}$ is the total density recorded over the region of the INMS measurement, including H$_2$, He, and CH$_4$ from INMS and H from the diffusion model. Thus we get the INMS volume mixing ratios of H$_2$, He, and CH$_4$ as follows
\begin{equation}\label{eqn:xINMS}
x_s^{\text{INMS}} = \frac{n_s^{\text{INMS}}}{n_{\text{tot}}^{\text{INMS}}}.
\end{equation}

\begin{figure}
\centering
\includegraphics[width=.45\textwidth]{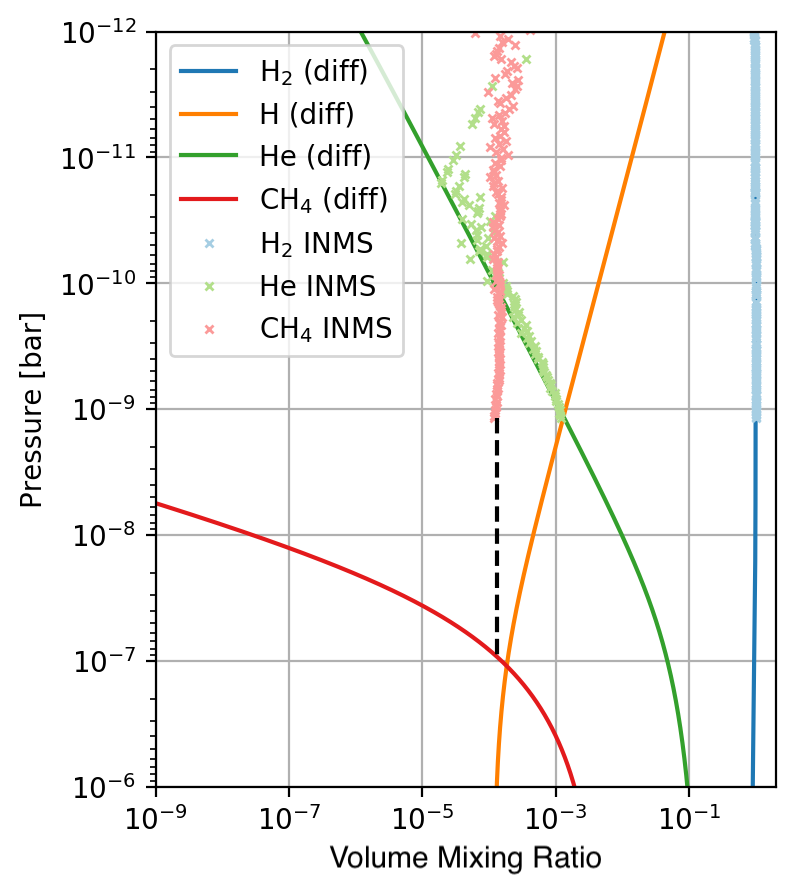}
\caption{Volume mixing ratios of the neutral species as a function of pressure from the diffusion model using composite temperature A (in solid lines), and INMS final plunge volume mixing ratios (crosses). The dashed black line shows the extension to higher pressures of the INMS CH$_4$ profile at a constant mixing ratio of $1.3\times 10^{-4}$.}
\label{fig:mixing_ratio_vs_p_INMS}
\end{figure}

The derived INMS volume mixing ratios from Equation~\ref{eqn:xINMS} are plotted with crosses in Fig.~\ref{fig:mixing_ratio_vs_p_INMS}, along with the mixing ratios from the diffusion model (solid lines). We adjusted the 1 bar value of the He volume mixing ratio so that the INMS data would match the diffusion model mixing ratios at pressures less than 1~nbar. Furthermore, as discussed in \citet{Yelle2018}, during the final plunge INMS measured an influx of CH$_4$, possibly from Saturn's rings. Therefore the INMS mixing ratios for CH$_4$ do not match those from the diffusion model. To construct an estimate of the CH$_4$ mixing ratio during the final plunge, we extend the INMS mixing ratio to higher pressures at a constant mixing ratio value of $1.3\times 10^{-4}$ (black dashed line in Fig.~\ref{fig:mixing_ratio_vs_p_INMS}), until we intersect the values from the diffusion model, which we use at pressures from $0.1~\mu$bar to 1~bar. Note that in reality there is likely a minimum in the CH$_4$ mixing ratio between the INMS and UVIS measurements, although this would not significantly change the results of this study.

\subsubsection*{Reconstructed neutral number densities}
\begin{figure*}
\centering
\includegraphics[width=.9\textwidth]{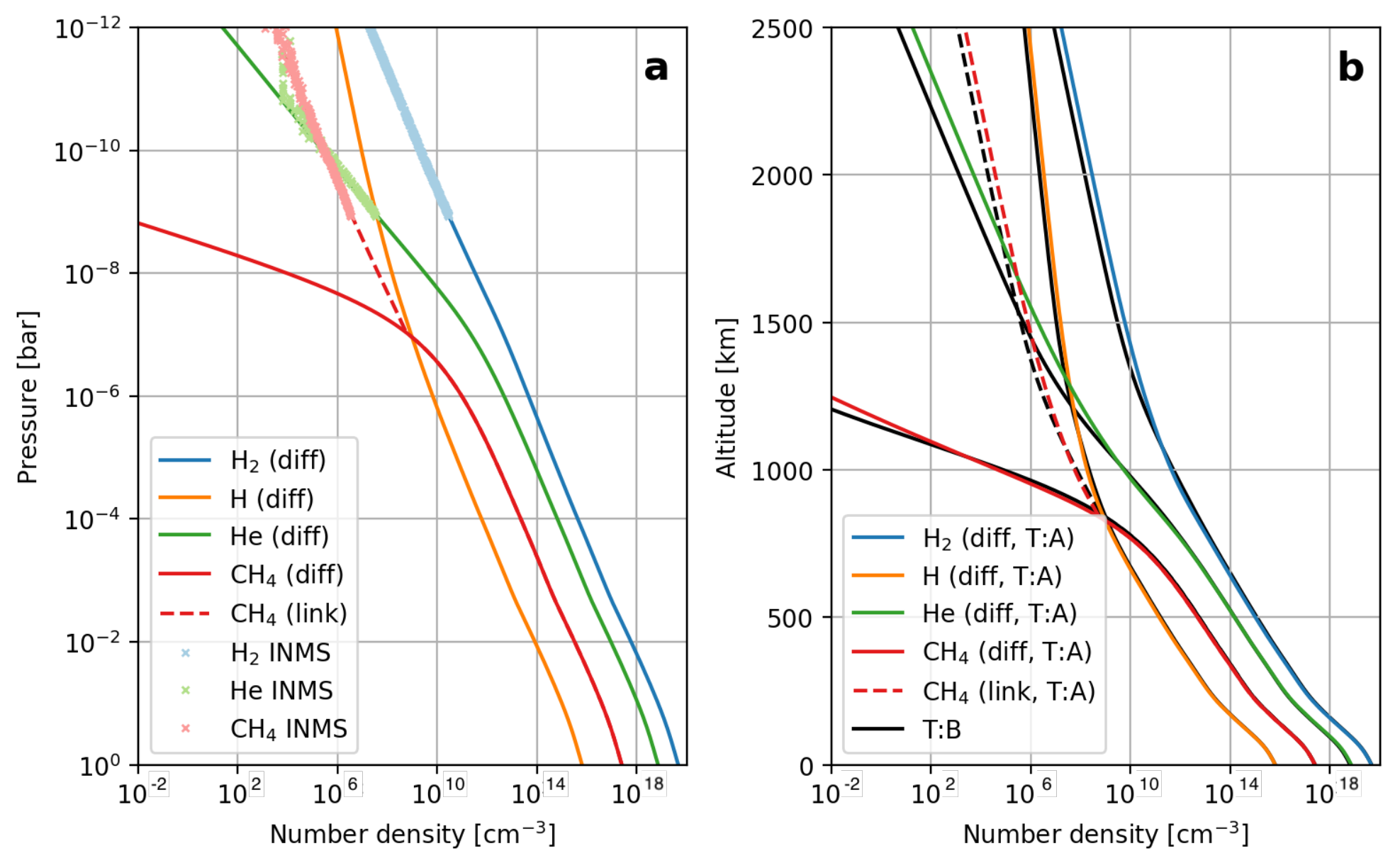}
\caption{Number density of the neutral species as a function of pressure (panel a), and as a function of altitude (panel b). Solid lines are densities obtained from the diffusion model, dashed lines show the link between the diffusion model and the INMS CH$_4$ profile using a constant mixing ratio of $1.3\times 10^{-4}$. In panel (a), the INMS final plunge densities are shown in crosses. In panel (b), a comparison is made between the densities obtained using temperature composite A (in coloured lines), and B (in black).}
\label{fig:n_vs_p_INMS}
\end{figure*}

We use the ideal gas law to obtain the total number density $n_{\text{tot}}$ from the total atmospheric pressure and reconstructed temperature profiles:
\begin{equation}
n_{\text{tot}} = \frac{p}{kT},
\end{equation}
where $k$ is the Boltzmann constant. Multiplying the volume mixing ratios by the total number density gives us the number density profiles for each neutral species, as shown in Fig.~\ref{fig:n_vs_p_INMS}. The density profiles as a function of pressure (Fig.~\ref{fig:n_vs_p_INMS}a) are independent of the temperature profile, however there is a difference in the densities above about 1100~km between profiles calculated using temperature composite A (coloured lines in Fig.~\ref{fig:n_vs_p_INMS}b), and composite B (black lines in Fig.~\ref{fig:n_vs_p_INMS}b).

\subsection{Solar flux}\label{sec:solar_flux}
We have access to solar spectral data at a range of different spectral resolutions, however none of these quite match the high resolution of the H$_2$ photo-absorption cross section model ($\Delta\lambda = 10^{-3}$~nm, see Sect.~\ref{sec:sigma_H2}). The coarsest resolution spectrum we consider is from TIMED/SEE at a wavelength resolution of $\Delta\lambda = 1$~nm \citep{Woods2005}. We also make use of a slightly higher resolution spectrum ($\Delta\lambda = 0.1$~nm) from the Whole Heliosphere Interval (WHI) quiet Sun campaign \citep{Woods2009,Chamberlin2009}. Our highest resolution spectrum is a quiet Sun reference spectrum from the SOHO/SUMER instrument at $\Delta\lambda = 4\times 10^{-3}$~nm \citep{Curdt2001}. Some characteristics of these datasets and the date of the observations used are presented in Table~\ref{tab:sol_spectra}.

\begin{table*}
\centering
\begin{threeparttable}
\caption{Solar flux data sources}
\begin{tabular}{lcccc}
\toprule
            & Wavelength & Sampling & Resolution of & Date of  \\
            & range$^*$      & resolution$^{\dagger}$ & spectrum used & observation \\
            & [nm]             & [nm]       & [nm]          &   \\
\midrule
TIMED/SEE$^1$  & 0.5 -- 152    & 0.4 -— 7   & 1             & 14 April 2008  \\
WHI$^2$        & 0.1 -- 152& 0.1 -— 7   & 0.1                    & 10 -- 16 April 2008 \\
SOHO/SUMER$^3$ & 67 -- 152     & 0.004      & 0.004                & 20 April 1997 \\
\bottomrule
\end{tabular}
\begin{tablenotes}
\small
\item Notes: $^*$Wavelength range used in this study, TIMED/SEE spectra extend to 190~nm and the complete WHI dataset extends to 2400~nm $^{\dagger}$Instrument resolution over the wavelength range used in this study.
\item Sources: $^1$\citet{Woods2005} $^2$\citet{Woods2009,Chamberlin2009} $^3$\citet{Curdt2001}.
\end{tablenotes}
\label{tab:sol_spectra}
\end{threeparttable}
\end{table*}

The final plunge on 15 September 2017 took place during quiet solar conditions, at solar minimum. We therefore use solar spectra during similar conditions, measured by F10.7 and Lyman $\alpha$ fluxes. The values of these reference fluxes on the day of the final plunge and on the dates each of the spectra we use in this study were recorded are shown in Table~\ref{tab:sol_conditions}.

\begin{table}
\centering
\begin{threeparttable}
\caption{Solar flux conditions}
\begin{tabular}{lcc}
\toprule
Date & Lyman $\alpha$ flux & F10.7 \\
     & [$10^{11}$~cm$^{-2}$~s$^{-1}$] & [sfu] \\
\midrule
20 April 1997 & 3.56 & 70.4 \\
14 April 2008 & 3.50 & 69.0 \\
15 September 2017 & 3.61 & 73.6 \\
\bottomrule
\end{tabular}
\label{tab:sol_conditions}
\end{threeparttable}
\end{table}

The WHI quiet Sun spectrum is a composite of observations over different wavelength bands. In the soft X-ray, EUV, and FUV (the wavelengths that are absorbed in upper planetary atmospheres), the WHI dataset is composed of measurements from the XPS instrument on TIMED/SEE between 0.1 -- 6.0~nm, a rocket measurement between 6.0 -- 105~nm, the EGS instrument on TIMED/SEE between 105 -- 116~nm, and a SORCE/SOLSTICE spectrum beyond 116~nm. The rocket was launched on 14 April 2008, and the TIMED/SEE and SORCE/SOLSTICE spectra that compose the WHI spectrum are averages over 10 -- 16 April 2008 (solar minimum, see Table~\ref{tab:sol_conditions}). 

The SUMER spectrum extends from 67 to 152~nm at a resolution of $4\times 10^{-3}$~nm (see Table~\ref{tab:sol_spectra}). We combine this spectrum either with a low resolution TIMED/SEE spectrum or the WHI spectrum in order to produce solar spectra over 0.1 to 152~nm. In addition, since the SUMER instrument only observes a small portion of the Sun at the centre of the solar disk, it measures spectral radiances. Thus we must convert these data to irradiances (a disk-integrated quantity). We do so by ensuring that the integrated SUMER spectrum matches the integrated flux from TIMED/SEE or WHI over $\lambda_1 =67$~nm to $\lambda_2 =152$~nm:
\begin{equation}\label{eqn:conversion_irrad}
I_{\lambda}^{\text{SUMER}} = \frac{\int_{\lambda_1}^{\lambda_2} \! I_{\lambda'}^{\text{SEE}} \, \mathrm{d}\lambda'}{\int_{\lambda_1}^{\lambda_2} \! L_{\lambda'}^{\text{SUMER}} \, \mathrm{d}\lambda'}  L_{\lambda}^{\text{SUMER}},
\end{equation}
where $I_{\lambda}$ are irradiances and $L_{\lambda}$ are radiances. This results in the integrated irradiance over 67 -- 152~nm from the TIMED/SEE spectrum and from the WHI spectrum being both the same, equal to 8.6~mW~m$^{-2}$.

Equation~\ref{eqn:conversion_irrad} provides only an approximation of solar irradiances. By scaling the SUMER radiances in this way, we are assuming that the radiance at disk centre is representative of that of the entire disk. This means that we neglect the contribution of active regions and coronal holes to the full disk irradiance. However, given that we consider a quiet Sun, these contributions should be small \citep{Schuhle1998}. A larger source of error is neglecting centre-to-limb variability, which can appear as either limb brightening or darkening, depending on the particular spectral line. An accurate conversion of the SUMER measurement to a disk-integrated spectrum is not trivial and is beyond the scope of this study.

Using the three sources of solar spectral data described above (and presented in Table~\ref{tab:sol_spectra}), we construct four solar spectra: one `low', one `mid', and two `high' resolution spectra. The data sources used in each of these cases are given in Table~\ref{tab:construced_spectra}. The low-resolution spectrum has a resolution of $\Delta\lambda=1$~nm, and is composed of TIMED/SEE between 0.1~nm and 67~nm, and the SUMER spectrum degraded in resolution to $\Delta\lambda=1$~nm between 67 and 152~nm. The mid-resolution spectrum has $\Delta\lambda=0.1$~nm and is composed of the WHI spectrum between 0.1 and 67~nm combined with the SUMER spectrum degraded in resolution to $\Delta\lambda=0.1$~nm between 67 and 152~nm. The two high-resolution spectra (labelled \#1 and \#2) are constructed using the SUMER spectrum at $\Delta\lambda=0.004$~nm between 67 -- 152~nm, combined with either TIMED/SEE at $\Delta\lambda=1$~nm (spectrum \#1) or the WHI spectrum at $\Delta\lambda=0.1$~nm (spectrum \#2) between 0.1 -- 67~nm.

\begin{table}
\centering
\begin{threeparttable}
\caption{Constructed solar spectra}
\begin{tabular}{l|c|c}
\toprule
  & \multicolumn{2}{c}{Wavelength range} \\
\hline
Label    & 0.1 -- 67~nm & 67 -- 152~nm \\
\hline
\multirow{2}{*}{Low res.} & TIMED/SEE & SOHO/SUMER \\
         &   $\Delta\lambda=1$~nm        & degraded to $\Delta\lambda=1$~nm \\   
\hline
\multirow{2}{*}{Mid res.} & WHI      & SOHO/SUMER \\
        &    $\Delta\lambda=0.1$~nm        & degraded to $\Delta\lambda=0.1$~nm \\
\hline
\multirow{2}{*}{High res.\ \#1} & TIMED/SEE & SOHO/SUMER \\
        &   $\Delta\lambda=1$~nm        & full resolution \\
        &                                                & ($\Delta\lambda=0.004$~nm) \\
\hline
\multirow{2}{*}{High res.\ \#2} & WHI & SOHO/SUMER \\
        &  $\Delta\lambda=0.1$~nm   & full resolution \\
        &                                             & ($\Delta\lambda=0.004$~nm) \\
\bottomrule
\end{tabular}
\label{tab:construced_spectra}
\end{threeparttable}
\end{table}

For our purposes, high-resolution spectra are only required at wavelengths where the chemical cross sections in our atmospheric model are highly structured. In the H$_2$, H, He, and CH$_4$ atmosphere that we consider, the H$_2$ photo-absorption cross section is very structured at $\lambda > 80.4$~nm, i.e.\ beyond the ionisation threshold of H$_2$ (see Sect.~\ref{sec:sigma_H2}). At wavelengths shorter than this value, the chemical cross sections vary smoothly. Hence, when constructing our `high-resolution' solar spectra, it is sufficient to use either a low-resolution TIMED/SEE observation (for case high res.\ \#1) or a mid-resolution WHI spectrum (for case high res.\ \#2) at shorter wavelengths. In addition, in order to capture differences in ionisation rates due only to the spectral resolution, and not to the precise spectral energy distribution, we rebin the SUMER spectrum to 1~nm resolution to use in the longer wavelength region of the low-resolution spectrum and we rebin SUMER to 0.1~nm resolution to use in the mid-resolution spectrum (see Table~\ref{tab:construced_spectra}).

\begin{figure*}
\centering
\includegraphics[width=1\textwidth]{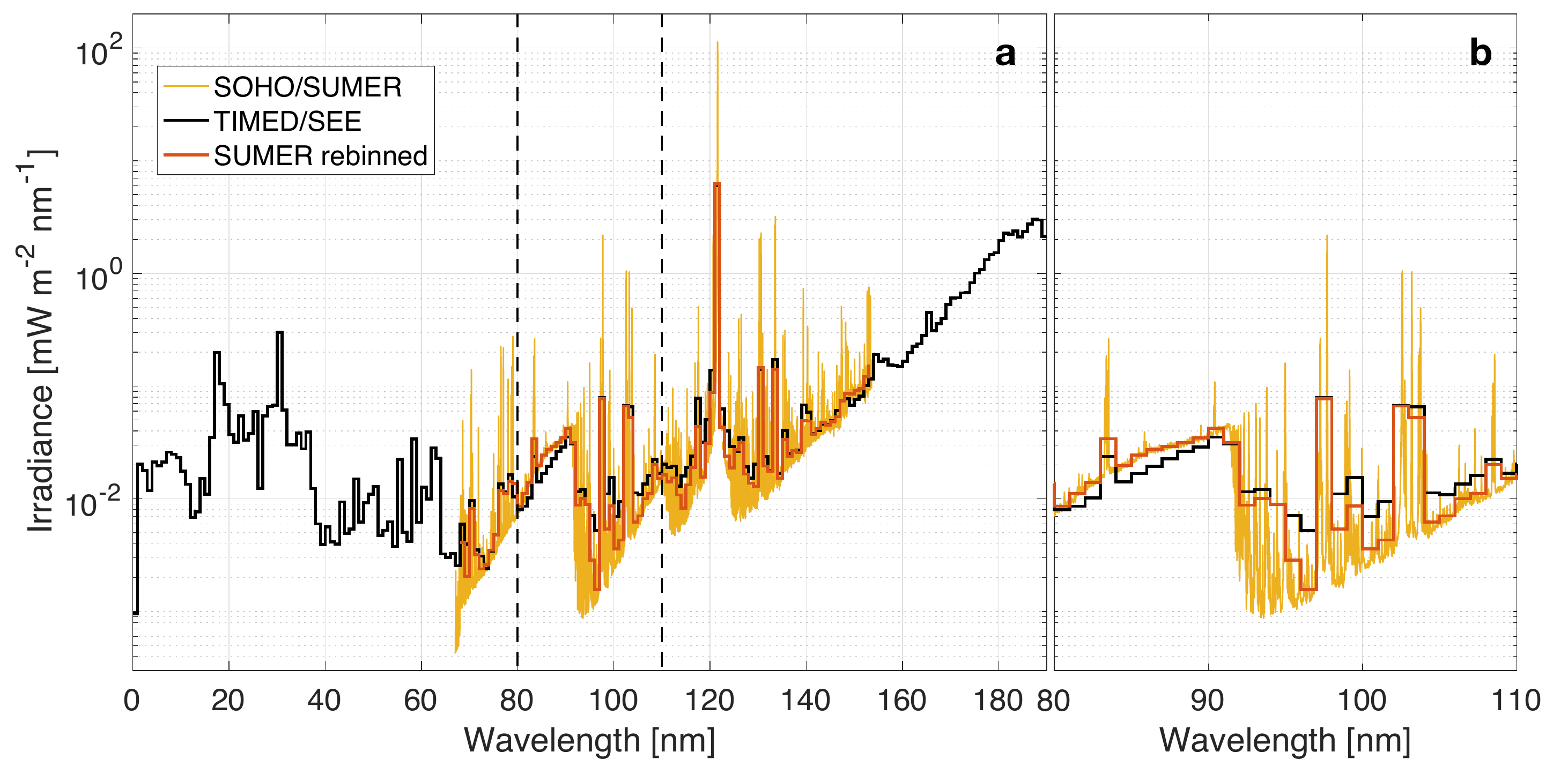}
\caption{Solar spectra from TIMED/SEE on 14 April 2008 (in black), and from SOHO/SUMER on 20 April 1997 (in yellow, and degraded to 1~nm resolution in red). Panel (b) is an enlargement of panel (a) between 80 and 110~nm (indicated by vertical dashed lines).}
\label{fig:spectrum_SEE_SUMER}
\end{figure*}

\begin{figure*}
\centering
\includegraphics[width=1\textwidth]{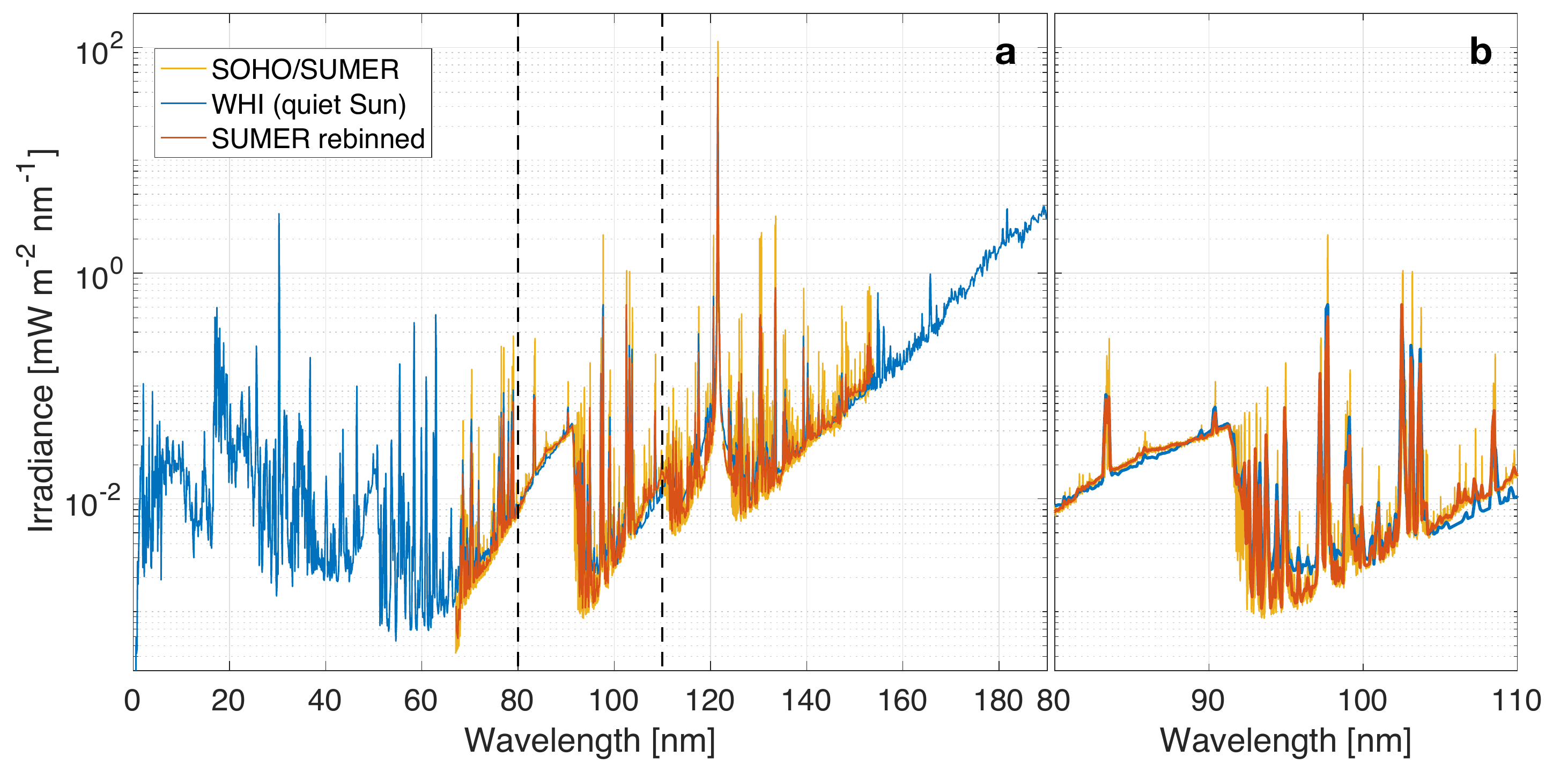}
\caption{Solar spectra from WHI quiet Sun campaign 10 -- 16 April 2008 (in blue), and from SOHO/SUMER on 20 April 1997 (in yellow, and degraded to 0.1~nm resolution in red). Panel (b) is an enlargement of panel (a) between 80 and 110~nm (indicated by vertical dashed lines).}
\label{fig:spectrum_WHI_SUMER}
\end{figure*}

The solar spectra are plotted in Figs.~\ref{fig:spectrum_SEE_SUMER} and \ref{fig:spectrum_WHI_SUMER}. Figure~\ref{fig:spectrum_SEE_SUMER} shows the TIMED/SEE spectrum (in black) along with the SOHO/SUMER spectrum at full resolution (in yellow), and the latter rebinned to the same resolution as TIMED/SEE, i.e.\ $\Delta\lambda=1$~nm (in red). Panel (b) is an enlargement of panel (a) focussing on the spectral region where the H$_2$ photo-absorption cross section is highly structured: from 80~nm to 110~nm. Beyond 110~nm, the cross section drops off at the atmospheric temperatures considered (see Sect.~\ref{sec:sigma_H2}). The rebinned SUMER spectrum allows a comparison with the SEE flux levels. We expect the fluxes to be similar since we have applied Equation~\ref{eqn:conversion_irrad} to obtain the SUMER spectral irradiance, setting the integrated flux between 67 and 152~nm to be identical between the two spectra. In general, there is a good agreement between the spectral shapes observed by SEE and SUMER, especially since these spectra are measured during different solar cycles (see Table~\ref{tab:sol_conditions}). Most importantly, the bins containing strong spectral lines match very well. The largest discrepancies occur in the bins showing lower fluxes, in particular near 95~nm, which has previously been noted by \citet{Curdt2001}. Overall we obtain a high coefficient of determination of $R^2=0.92$, when performing a linear fit between the TIMED/SEE spectrum between 80~nm and 110~nm and the SUMER spectrum rebinned to 1~nm resolution.

Figure~\ref{fig:spectrum_WHI_SUMER} shows the WHI quiet Sun spectrum from 0.1 to 190~nm in blue upon which the full-resolution SUMER spectrum is superposed in yellow. The SUMER spectrum rebinned to match the resolution of WHI ($\Delta\lambda=0.1$~nm) is plotted in red. Once again, the two observations are very close: $R^2=0.82$ over the wavelength range 80~nm to 110~nm. Certain wavelength regions of the SUMER spectrum (e.g. the Lyman continuum at $\lambda<91.2$~nm) match better with the WHI spectrum than with that from SEE, which could be due to instrumental effects caused by the degradation of the SEE sensor. We do not expect exact agreement between the different spectra given solar variability, instrumental noise and that SUMER only measures a portion of the solar disk.

Regions where the SEE or WHI spectra are slightly higher than the SUMER spectrum can be explained by the fact that most solar emission lines at these wavelengths undergo limb brightening \citep{Wilhelm1998,Schuhle1998}. Since the SUMER observation only includes the disk centre, our SUMER irradiances are missing this component. Despite this fact, the SEE, WHI, and SUMER spectra agree very well with each other, and the SUMER spectrum processed with Equation~\ref{eqn:conversion_irrad} is sufficient for the purposes of this study.

\subsection{High-resolution H$_2$ photo-absorption cross section}\label{sec:sigma_H2}
The photo-absorption cross section of molecular hydrogen is highly structured at wavelengths longer than the ionisation threshold ($\lambda_{\text{th,H}_2}=80.4$~nm). It is made up of very narrow absorption lines composing the Lyman, Werner, and Rydberg bands \citep{Abgrall1993a,Abgrall1993,Abgrall2000}. These lines result in the absorption of solar radiation by the H$_2$ molecule over an extended layer of atmosphere that can only be modelled by including a high-resolution H$_2$ cross section. Previous studies have found that absorption in the Lyman, Werner, and Rydberg bands can produce a layer of hydrocarbon ions in the lower ionosphere of Jupiter \citep{Kim1994} and Saturn \citep{Kim2014}.

\begin{figure*}
\centering
\includegraphics[width=1\textwidth]{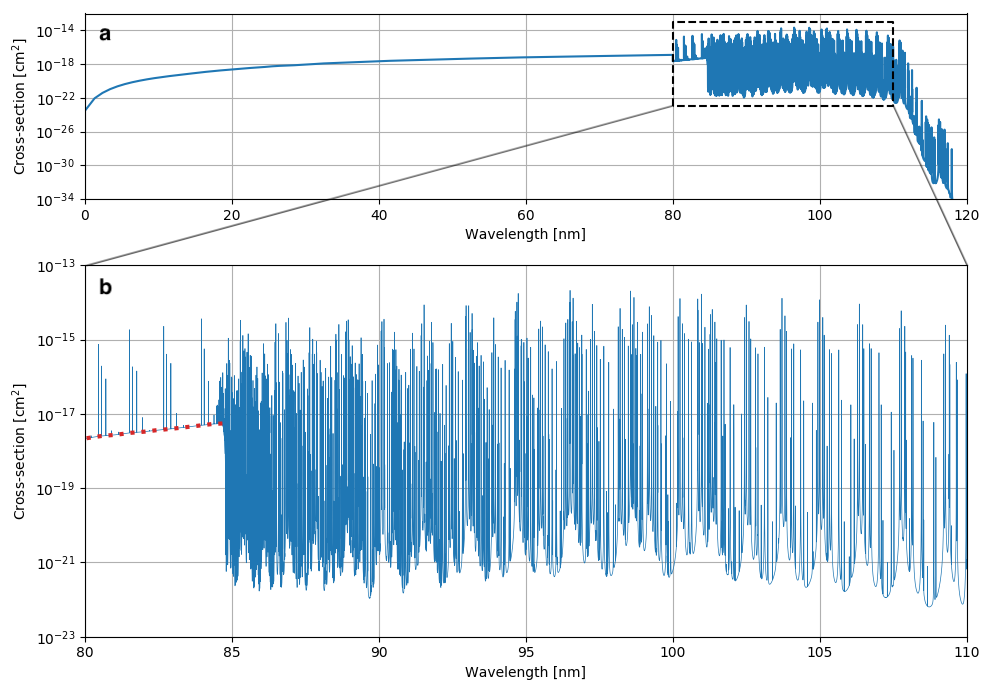}
\caption{Panel a shows the molecular hydrogen photo-absorption cross section, using measurements by \citet{Backx1976} at $\lambda<80.4$, and high-resolution ($\Delta\lambda=10^{-3}$~nm) calculations from \citet{Yelle1993} at a temperature of 250~K at $\lambda>80.4$. Between 80.4 and 84.6~nm, the photo-dissociation cross section measured by \citet{Dalgarno1969} is included. Panel b is an enlargement of panel a, focussing on the highly structured region beyond the H$_2$ ionisation threshold. The contribution from the \citet{Dalgarno1969} photo-dissociation cross section is highlighted in a red dotted line.}
\label{fig:sigH2}
\end{figure*}

For the H$_2$ photo-absorption cross section used in this study, we take a combination of the \citet{Backx1976} low-resolution ($\Delta\lambda\sim 1$~nm) H$_2$ photo-absorption cross section for wavelengths below the ionisation threshold (80.4~nm) and temperature-dependent high-resolution ($\Delta\lambda=10^{-3}$~nm) calculations by \citet{Yelle1993} at longer wavelengths. In addition, the H$_2$ photo-dissociation cross section measured by \citet{Dalgarno1969} was added to the \citet{Yelle1993} calculations for wavelengths between 80.4~nm and 84.6~nm where this process was missing. The cross section is plotted in Fig.~\ref{fig:sigH2}, and is provided as a downloadable dataset \citep{Chadney2021}.

Although the high-resolution H$_2$ cross section is dependent on temperature, in practice, the temperature matters little: there is a maximum difference of 15\% in some peak ionisation rates between model runs using \citet{Yelle1993} cross sections determined at 150~K and at 350~K (the range of thermospheric temperatures, see Fig.\ref{fig:Tsource}). However, this may be because even our high resolution solar spectrum from SUMER ($\Delta\lambda=4\times 10^{-3}$~nm) has a wavelength resolution that is too coarse. The effect of the temperature of the H$_2$ cross section on ionisation rates might be more significant if the calculations were done using a solar spectrum with a resolution closer to $\Delta\lambda=1\times 10^{-3}$~nm. Nevertheless, in our case the cross section temperature is not important, and in all the following calculations in this paper we use an H$_2$ high-resolution cross section from \citet{Yelle1993} at a temperature of 250~K; this value was chosen as mid way between the temperature at the bottom of the thermosphere and the exospheric temperature.

\section{Energy deposition model}\label{sec:ionospheric_model}
The energy deposition model developed for this study calculates a set of ionisation and photo-dissociation rates. The neutral species H$_2$, H, He, and CH$_4$ (profiles of which are determined in Sect.~\ref{sec:neutral_atm}) are ionised through photo-ionisation and electron-impact ionisation. The former is obtained by solving the Beer-Lambert law. We included the latter by using a suprathermal electron transport model that is based on the solution to the Boltzmann equation with transport, angular scattering, and energy degradation of photo-electrons and their secondaries taken into account \citep{Galand2009}. The incident source of energy is the solar spectrum derived in Sect.~\ref{sec:solar_flux}. Since we are modelling the equatorial atmosphere, we do not include electron precipitation from the magnetosphere.

\begin{figure*}
\centering
\includegraphics[width=1\textwidth]{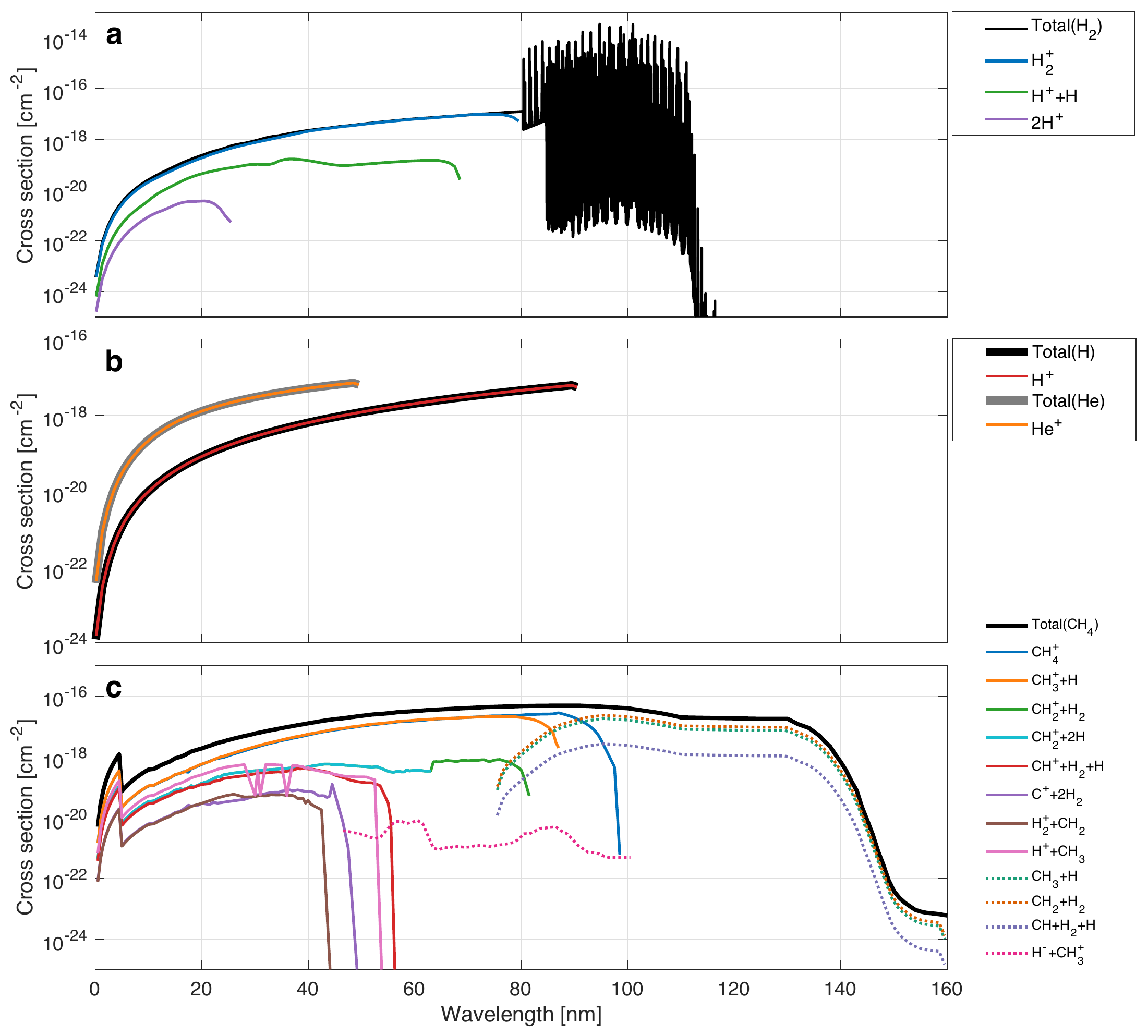}
\caption{Photo-absorption (black and grey lines), photo-ionisation (solid coloured lines), and photo-dissociation (dotted coloured lines) cross sections for the four neutral species H$_2$ (panel a), H and He (panel b), and CH$_4$ (panel c). In panel b the curves showing total H and He photo-absorption cross sections are respectively indistinguishable from the H$^+$ and He$^+$ photo-ionisation cross sections.}
\label{fig:cross_sections}
\end{figure*}

The photo-ionisation, electron-impact ionisation, and photo-dissociation reactions included are shown in Table~\ref{tab:reactionsiono} and the cross sections associated with each of these reactions are plotted in Fig.~\ref{fig:cross_sections}. For a more detailed description of the energy deposition model, see \citet{Chadney2016}.

\begin{table*}
\centering
\small
\caption{Chemical reactions used in the energy deposition model.}
\begin{tabular}{llcl}
\toprule
\# & Reaction & Reference & \\
\midrule
 & Photo-ionisation: & & \\
1 & $\text{H}_2 + h\nu \rightarrow \text{H}_2^+ + e^-$ & \multicolumn{2}{l}{\citet{Backx1976,Kossmann1989},} \\
  &													   & \multicolumn{2}{l}{\citet{Chung1993,Yan1998}} \\
2 & $\text{H}_2 + h\nu \rightarrow \text{H}^+ + \text{H} + e^-$ & \multicolumn{2}{l}{\citet{Chung1993} and 2H$^+$ references} \\
3 & $\text{H}_2 + h\nu \rightarrow 2\text{H}^+ + 2e^-$ & \multicolumn{2}{l}{\citet{Dujardin1987,Kossmann1989a},} \\
  &													   & \multicolumn{2}{l}{\citet{Yan1998}} \\
4 & $\text{H} + h\nu \rightarrow \text{H}^+ + e^-$ & \multicolumn{2}{l}{\citet{Verner1996}} \\
5 & $\text{He} + h\nu \rightarrow \text{He}^+ + e^-$ & \multicolumn{2}{l}{\citet{Verner1996}} \\
6 & $\text{CH}_4 + h\nu \rightarrow \text{CH}_4^+ + e^-$ & \multirow{8}{*}{\hspace{-1em}$\left.\begin{array}{l}
                \\
                \\
                \\
                \\
                \\
                \\
                \\
                \\
                \end{array}\right\rbrace$\,\,\citet{Samson1989,Schunk2000}} \\
7 & $\text{CH}_4 + h\nu \rightarrow \text{CH}_3^+ + \text{H} + e^-$ &  \\
8 & $\text{CH}_4 + h\nu \rightarrow \text{CH}_2^+ + \text{H}_2 + e^-$ &  \\
9 & $\text{CH}_4 + h\nu \rightarrow \text{CH}_2^+ + 2\text{H} + e^-$ &  \\
10 & $\text{CH}_4 + h\nu \rightarrow \text{CH}^+ + \text{H}_2 + \text{H} + e^-$ & \\
11 & $\text{CH}_4 + h\nu \rightarrow \text{C}^+ + 2\text{H}_2 + e^-$ &  \\
12 & $\text{CH}_4 + h\nu \rightarrow \text{H}_2^+ + \text{CH}_2 + e^-$ &  \\
13 & $\text{CH}_4 + h\nu \rightarrow \text{H}^+ + \text{CH}_3 + e^-$ &  \\
 & & & \\
 & Electron-impact ionisation: & & \\
14 & $\text{H}_2 + e^- \rightarrow \text{H}_2^+ + e^- + e^-$ & \multicolumn{2}{|l}{\citet{VanWingerden1980,Ajello1991},} \\
15 & $\text{H}_2 + e^- \rightarrow \text{H}^+ + \text{H} + e^- + e^-$ & \multicolumn{2}{|l}{\citet{Jain1992,Straub1996},} \\
16 & $\text{H}_2 + e^- \rightarrow 2\text{H}^+ + 2e^- + e^-$ & \multicolumn{2}{|l}{\citet{Liu1998,Brunger2002}} \\
17 & $\text{H} + e^- \rightarrow \text{H}^+ + e^- + e^-$ & \multicolumn{2}{l}{\citet{Brackmann1958,Burke1962},} \\
  &														& \multicolumn{2}{l}{\citet{Bray1991,MAYOL1997},} \\
  &														& \multicolumn{2}{l}{\citet{Stone2002,Bartlett2004}} \\
18 & $\text{He} + e^- \rightarrow \text{He}^+ + e^- + e^-$ & \multicolumn{2}{l}{\citet{LaBahn1970,MAYOL1997},} \\
   &													   & \multicolumn{2}{l}{\citet{Stone2002,Bartlett2004}} \\
19 & $\text{CH}_4 + e^- \rightarrow \text{CH}_4^+ + e^- + e^-$ & \multirow{8}{*}{\hspace{-1em}$\left.\begin{array}{l}
                \\
                \\
                \\
                \\
                \\
                \\
                \\
                \\
                \end{array}\right\rbrace$\,\,\citet{Davies1989,Liu2006}} \\
20 & $\text{CH}_4 + e^- \rightarrow \text{CH}_3^+ + \text{H} + e^- + e^-$ &  \\
21 & $\text{CH}_4 + e^- \rightarrow \text{CH}_2^+ + \text{H}_2 + e^- + e^-$ &  \\
22 & $\text{CH}_4 + e^- \rightarrow \text{CH}_2^+ + 2\text{H} + e^- + e^-$ &  \\
23 & $\text{CH}_4 + e^- \rightarrow \text{CH}^+ + \text{H}_2 + \text{H} + e^- + e^-$ &  \\
24 & $\text{CH}_4 + e^- \rightarrow \text{C}^+ + 2\text{H}_2 + e^- + e^-$ &  \\
25 & $\text{CH}_4 + e^- \rightarrow \text{H}_2^+ + \text{CH}_2 + e^- + e^-$ &  \\
26 & $\text{CH}_4 + e^- \rightarrow \text{H}^+ + \text{CH}_3 + e^- + e^-$ &  \\
& & & \\
& Photo-dissociation: & & \\
27 & $\text{CH}_4 + h\nu \rightarrow \text{CH}_3 + \text{H}$ & \multirow{4}{*}{$\left.\begin{array}{l}
                \\
                \\
                \\
                \\
                \end{array}\right\rbrace$\,\,\citet{Lavvas2011}, based upon \citet{Wang2000}} \\
28 & $\text{CH}_4 + h\nu \rightarrow \text{CH}_2 + \text{H}_2$ &  \\
29 & $\text{CH}_4 + h\nu \rightarrow \text{CH} + \text{H}_2 + \text{H}$ &  \\
30 & $\text{CH}_4 + h\nu \rightarrow \text{H}^- + \text{CH}_3^+$ &  \\
\bottomrule
\end{tabular}
\label{tab:reactionsiono}
\end{table*}

\section{Results and Discussion} \label{sec:results}

\subsection{Ionisation rates} \label{sec:ionisation_rates}
\begin{figure*}
\centering
\includegraphics[width=0.8\textwidth]{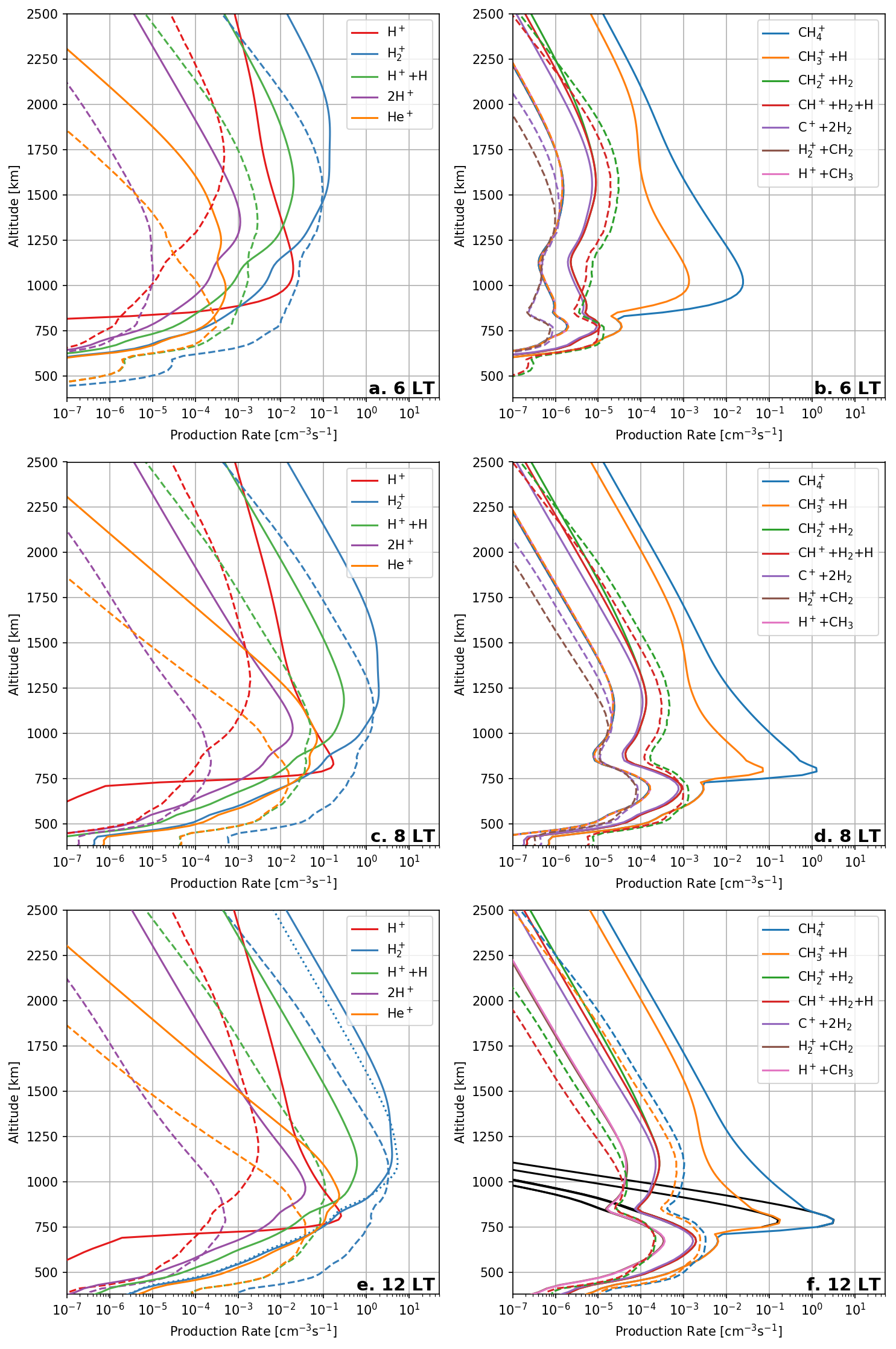}
\caption{Photo-ionisation rates (solid coloured lines) and electron-impact ionisation rates (dashed lines) during Grand Finale conditions calculated using the high resolution solar spectrum \#1 and the high resolution H$_2$ photo-absorption cross section. Panels a and b show profiles calculated at 6~LT, panels c and d, at 8~LT, and panels e and f, at 12~LT. Panel f also contains in solid black curves the CH$_4$ photo-ionisation profiles obtained if there were no inflow of CH$_4$ (i.e.\, using CH$_4$ densities from the diffusion model plotted in the solid red curve in Fig.~\ref{fig:n_vs_p_INMS} and discussed in Sect.~\ref{sec:neutral_atm}). All profiles are calculated using a neutral atmosphere derived with temperature composite A (see Sect.~\ref{sec:neutral_atm}), apart from the blue dotted curve in panel e, which shows the effect of using temperature composite B on the H$_2^+$ photo-ionisation rate.}
\label{fig:Pi_vs_z}
\end{figure*}

Ionisation rates through the reactions listed in Table~\ref{tab:reactionsiono} are plotted as a function of altitude in Fig.~\ref{fig:Pi_vs_z}, where the solid lines are the rates of photo-ionisation reactions and the dashed lines are the rates of the corresponding electron-impact ionisation reactions. The reaction rate profiles are shown at three different local times: 6~LT (panels a,b), 8~LT (panels c,d), and 12~LT (panels e,f).  In panels a, c, and e we have plotted the rates of reactions forming H$_2^+$, H$^+$, and He$^+$ through reactions \#~1 to 5 and \#~14 to 18 (from Table~\ref{tab:reactionsiono}) for photo-ionisation and electron-impact ionisation, respectively. Panels b, d, and f show production rates from the ionisation of CH$_4$ leading to CH$_4^+$, CH$_3^+$, CH$_2^+$, CH$^+$, C$^+$, H$_2^+$, and H$^+$ by reactions \#~6 to 13 for photo-ionisation and \#~19 to 26 for electron-impact ionisation. The neutral atmospheric profiles are not expected to be significantly affected by local time for a fast rotating planet. It is hence legitimate to use them over the range of local times in order to assess the ionisation and photo-dissociation rates at different LTs.

Photo-ionisation is the main ionisation process in the upper part of the ionosphere, above about 1000~km at 12 LT. At lower altitudes, the energy deposition of high-energy solar soft X-ray radiation ($\lambda < 20$~nm) results in large quantities of energetic secondary electrons allowing electron-impact ionisation to dominate, confirming earlier findings \citep[e.g.,][]{Kim1994,Galand2009,Kim2014}

At all local times, the main ion formed in the upper ionosphere is H$_2^+$. In terms of photo-ionisation H$_2^+$ is the dominant ion produced above 850~km at 12~LT at which time its photo-ionisation production rate displays a broad peak between 1000 and 1500~km at 4.2~cm$^{-3}$~s$^{-1}$. This peak is due to photons from the strong solar He II line at 30.4~nm. Although it is the main ion formed, previous modelling \citep[e.g.,][]{Moore2004,Kim2014} shows that the H$_2^+$ ion is efficiently converted into H$_3^+$ at high altitudes by reaction with abundant molecular H$_2$ (see Fig.~\ref{fig:n_vs_p_INMS}), which yields IR thermal emissions \citep{ODonoghue2013}.

The highly structured nature of the H$_2$ photo-absorption cross section at wavelengths beyond the H$_2$ ionisation threshold allows for low-energy photons (with wavelengths 84.6 - 120~nm) to penetrate down to altitudes as low as 800~km at 6~LT and $\sim700$~km at 12~LT if their energy falls within the wings of the very narrow lines that constitute the Lyman, Werner and Rydberg bands of the cross section. The result is a low-altitude narrow peak in the photo-ionisation rate profiles of atomic H to form H$^+$, and of CH$_4$ to form CH$_4^+$ and CH$_3^+$. These peaks are located between 750 and 850~km altitude at 12 LT, and stand out as being distinct in shape to the other production rate profiles in Fig.~\ref{fig:Pi_vs_z}. They result from ionisation associated with energy thresholds beyond 80~nm and are therefore sensitive to the highly structured H$_2$ photoabsorption cross section in this spectral region. At its peak near 800~km altitude, the rate of reaction \#~6 producing CH$_4^+$ reaches 3.2~cm$^{-3}$~s$^{-1}$ at 12 LT, making it the dominant reaction at this altitude.

As shown in Fig.~\ref{fig:Pi_vs_z}(b,d,f), the production profiles of other hydrocarbon ions (CH$_2^+$, CH$^+$, C$^+$), and H$_2^+$ and H$^+$ from the ionisation of CH$_4$ do not display the strong peaks of the CH$_4^+$ and CH$_3^+$ rates. Indeed, apart from the reactions producing CH$_4^+$ and CH$_3^+$, the other photo-ionisation reactions of CH$_4$ included in this study have ionisation threshold wavelengths that are too low to be affected by low-altitude absorption of photons in the structured region of the H$_2$ cross-section (see Fig.~\ref{fig:cross_sections}). Instead the production rate profiles of these ions have two peaks: a low-altitude peak in the region where solar soft X-ray photons are absorbed, and a another, broader peak at higher altitudes caused by solar EUV photons interacting with the high quantities of neutral CH$_4$ discovered to be present at high altitudes (see Fig.~\ref{fig:n_vs_p_INMS}).

The solar zenith angle decreases as local time moves towards noon. At higher zenith angles, a more extended column of atmosphere is required to achieve a given amount of absorption. Hence, the production peaks at 6 LT (panels a and b of Fig.~\ref{fig:Pi_vs_z}) and at 8 LT (panels c, d) are broadened, shifted to higher altitudes, and less intense compared to 12 LT (panels e, f). In addition, the altitude below which electron-impact ionisation dominates over photo-ionisation increases with solar zenith angle as soft X-ray and energetic EUV photons are absorbed at higher altitudes: at 12 LT, electron-impact ionisation dominates below 750~km, whereas at 6 LT, the threshold altitude is near 900~km. Photo-ionisation and electron-impact ionisation are of similar magnitude between about 750~km and 1000~km at 12~LT, and between about 900~km and 1500~km at 6~LT.

In Sect.~\ref{sec:neutral_atm}, we derived neutral atmospheres based on two different temperatures profiles, composites A and B, each constructed using a different method to connect the INMS final plunge measurements to previous CIRS and UVIS observations. The photo-ionisation rate profiles shown in solid lines in Fig.~\ref{fig:Pi_vs_z} are produced using the neutral atmosphere derived from temperature composite A. To show the effect of the two different temperature profiles on the ionisation rates, in panel e of Fig.~\ref{fig:Pi_vs_z} we have additionally plotted the photo-ionisation rate profile of H$_2$ to form H$_2^+$ obtained using a neutral atmosphere derived from temperature composite B; this is shown in a dotted blue line. Differences between the two production rate profiles are seen at higher altitudes (where the temperature profiles differ). A neutral atmosphere derived from temperature composite B results in a peak H$_2^+$ photo-ionisation rate that is higher by a factor of 1.5, compared to calculations making use of a neutral atmosphere derived from temperature composite A. At altitudes above about 1700~km, H$_2^+$ photo-ionisation rates are lower by a factor of 2 when using temperature composite B, compared to A.

\subsection{H$_2^+$ number density}
\begin{figure*}
\centering
\includegraphics[width=1\textwidth]{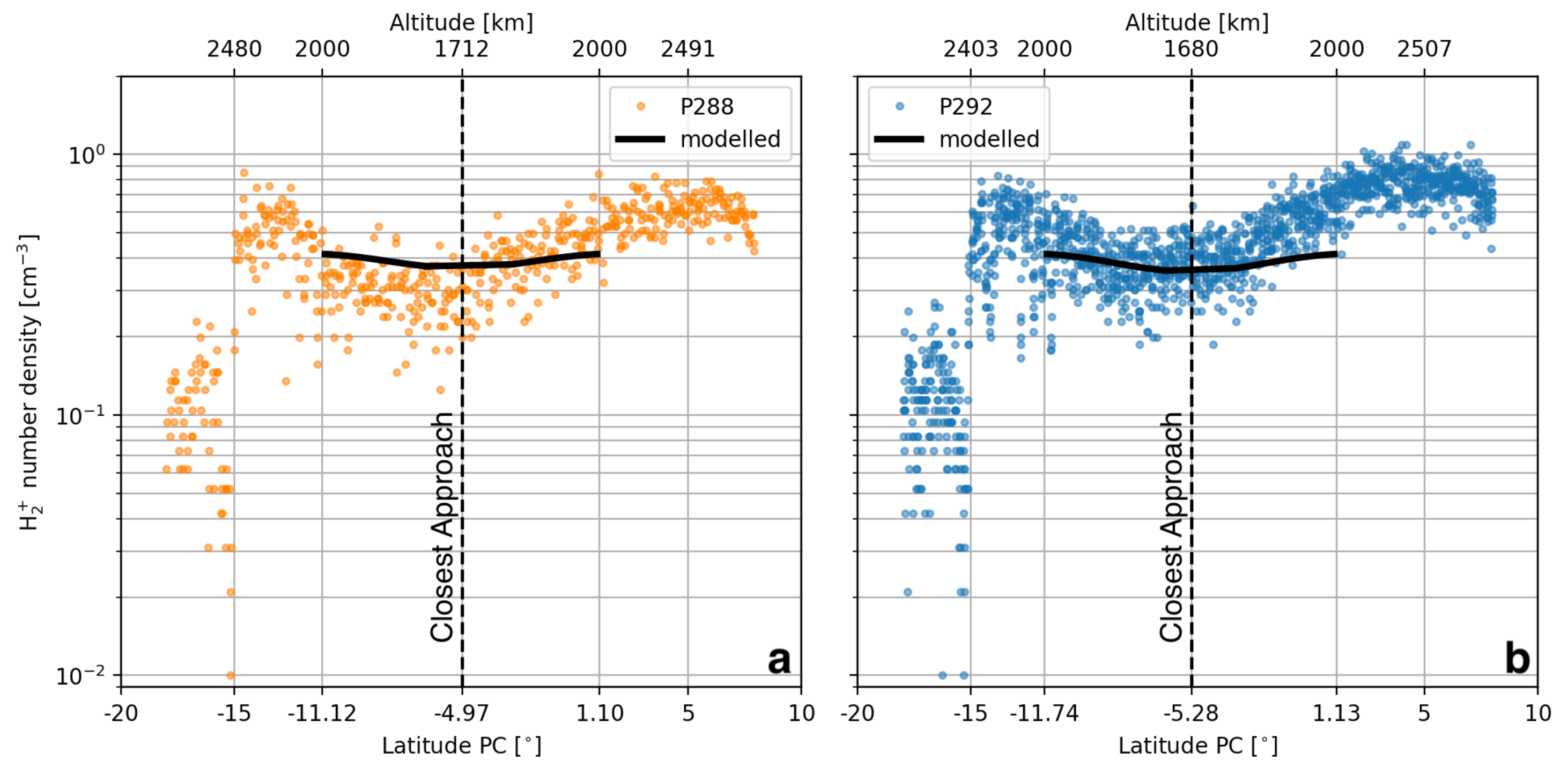}
\caption{Comparison of modelled H$_2^+$ number densities (black solid curve) with those measured by INMS during proximal orbit 288 (orange points, panel a) and orbit 292 (blue points, panel b), as a function of planetocentric latitude. The spacecraft closest approach (which took place near noon local time for both orbits) is shown with a vertical dashed line. Corresponding altitude values along the trajectory are also shown. Note that while the latitude scale is linear, this is not the case for the altitudes. INMS measurements are from \citet{Moore2018}. The modelled H$_2^+$ number densities presented here are determined at a solar zenith angle of 27$^{\circ}$.}
\label{fig:nH2p_vs_lat}
\end{figure*}

In order to validate our results, we compute the number density of the H$_2^+$ ion to compare with INMS measurements taken during Cassini's proximal orbits. The INMS instrument sampled Saturn's ionosphere down to altitudes close to 1700~km during the closest approach of orbits P288 and P292, which occurred on 14 August 2017 and 9 September 2017, respectively \citep{Moore2018,Waite2018,Cravens2019}. The measured H$_2^+$ densities along the trajectory of these orbits are shown in Fig.~\ref{fig:nH2p_vs_lat}, in orange points for P288 (panel a) and blue points for P292 (panel b). The measured densities reach a local minimum of about 0.2 - 0.5~cm$^{-3}$ at the altitude of closest approach. The sharp drop in density at latitudes below -15$^{\circ}$ is due to shadowing from the rings \citep{Moore2018}.

H$_2^+$ is the ion produced in the largest quantities above at least 1000~km in Saturn's upper atmosphere (see Fig.~\ref{fig:Pi_vs_z}). This ion reacts most efficiently with H$_2$ through proton transfer, which rapidly converts much of the population of H$_2^+$ into H$_3^+$:
\begin{equation}
\text{H}_2^+ + \text{H}_2 \xrightarrow{\text{k}} \text{H}_3^+ + \text{H},
\end{equation}
with a reaction rate of $k=2\times 10^{-9}$~cm$^3$s$^{-1}$ \citep{Theard1974}.

Below about 2000~km, chemical loss timescales are significantly lower than transport timescales and photochemical equilibrium is valid. There is a balance between the H$_2^+$ production rate ($P_{\text{H}_2^+}(z) = \nu_{\text{H}_2\rightarrow\text{H}_2^+}(z)\,n_{\text{H}_2}(z)$, where $\nu$ is the ionisation frequency) and its chemical loss: $L_{\text{H}_2^+}(z) = k\,n_{\text{H}_2^+}(z)\,n_{\text{H}_2}(z)$. Therefore we can determine the density of H$_2^+$ from our production rates (see Sect.~\ref{sec:ionisation_rates}) and the density of neutral H$_2$ (determined in Sect.~\ref{sec:neutral_atm}) according to the following:
\begin{equation} \label{eqn:nH2p}
n_{\text{H}_2^+}(z) = \frac{P_{\text{H}_2^+}(z)}{k\,n_{\text{H}_2}(z)}.
\end{equation}

Making use of Equation~\ref{eqn:nH2p}, we obtain the modelled H$_2^+$ number densities shown in the black solid curves in Fig.~\ref{fig:nH2p_vs_lat}, determined at noon local time. We take into account production of $\text{H}_2^+$ by photo-ionisation (reaction 1, Table~\ref{tab:reactionsiono}), and electron-impact ionisation due to photo-electrons (reaction 14, Table~\ref{tab:reactionsiono}). Electron-impact ionisation is responsible for about 10\% of the $\text{H}_2^+$ number density at these altitudes. The modelled density values shown in Fig.~\ref{fig:nH2p_vs_lat} are calculated at points along the INMS trajectory where the spacecraft's altitude is less than 2000~km, to ensure the assumption of photochemical equilibrium used in the above calculation is valid. We have also checked that our modelled density profile is not too sensitive to variations in solar zenith angle over these two periods.

Keeping in mind that the solar flux is our estimate at Saturn for quiet solar activity (see Sect.~\ref{sec:solar_flux}) without further adjustment and our atmospheric model was derived from the final plunge (on 15 September 2017, see Sect.~\ref{sec:neutral_atm}), the modelled and INMS $\text{H}_2^+$ number densities agree very well. In particular, at the closest approach altitude of 1700~km, we obtain a modelled value of $n_{\text{H}_2^+}=0.36$~cm$^{-3}$, which is consistent with the measurements around 0.2 - 0.5~cm$^{-3}$ from INMS. This calculated density corresponds to an ionisation frequency of H$_2$ producing H$_2^+$ of $\nu_{\text{H}_2\rightarrow\text{H}_2^+}=0.72\times 10^{-9}$~s$^{-1}$ ($\nu_{\text{H}_2\rightarrow\text{H}_2^+}(z)=k\,n_{\text{H}_2^+}(z)$). This value is in agreement with $(0.7\pm0.3)\times 10^{-9}$~s$^{-1}$, values derived by \citet{Cravens2019} from INMS measurements during orbit P288 at latitudes between -2$^{\circ}$ and -10$^{\circ}$ around the spacecraft closest approach.

\subsection{CH$_4$ photo-dissociation rates}
\begin{figure}
\centering
\includegraphics[width=0.5\textwidth]{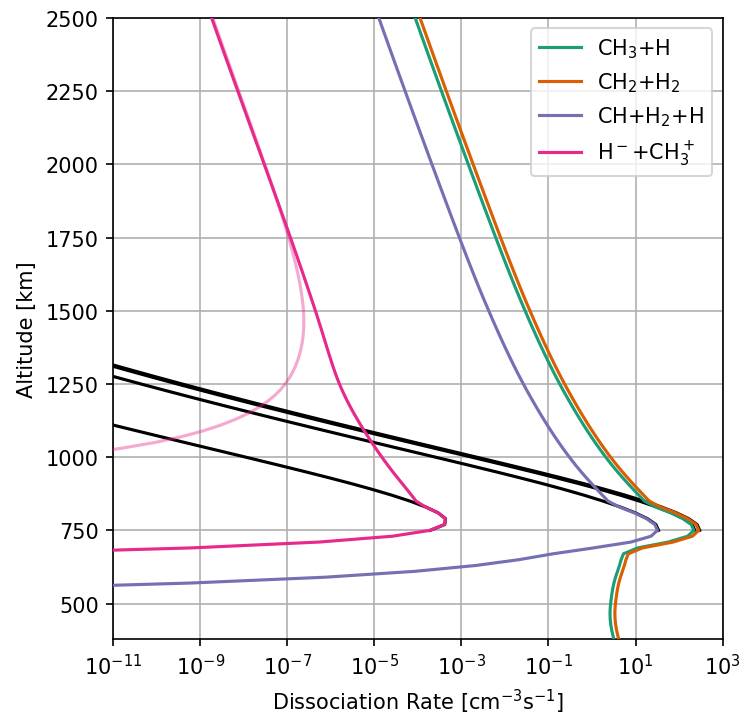}
\caption{CH$_4$ photo-dissociation rates during Grand Finale conditions at 12 LT calculated using the high resolution solar spectrum. The coloured curves show photo-dissociation rates including the inflow of CH$_4$ measured by INMS, whereas the black curves represent profiles obtained if there were no inflow of CH$_4$ (using the diffusion model discussed in Sect.~\ref{sec:neutral_atm} for the neutral CH$_4$ density profile). Only the reaction leading to H$^-$+CH$_3^+$ is sensitive to the spectral resolution of the H$_2$ photo-absorption cross section. The profile calculated using the high resolution H$_2$ cross section is shown in dark pink, and in light pink is the resulting profile when the low resolution H$_2$ cross section is used.}
\label{fig:Pd_vs_z}
\end{figure}

The dissociation rate profiles of CH$_4$ with altitude at 12 LT by reactions \#~27 to \#~30 (see Table~\ref{tab:reactionsiono}) are shown in Fig.~\ref{fig:Pd_vs_z}. The production rates of CH$_3$ (reaction \#~27), CH$_2$ (reaction \#~28), and CH (reaction \#~29), plotted respectively in green, orange, and purple, have the same dependence on altitude above the peak, since the shape of the cross sections of each of these reactions is identical (see Fig.~\ref{fig:cross_sections}c). The difference at low altitudes is due to different threshold wavelengths for the photons taking part in these reactions. The production peak is located at an altitude of 750~km and reaches 220~cm$^{-3}$~s$^{-1}$ for the production of CH$_3$, 280~cm$^{-3}$~s$^{-1}$ for CH$_2$, and 30~cm$^{-3}$~s$^{-1}$ for CH. These three reactions are driven by Lyman $\alpha$ photons and their rate profiles are not sensitive to the highly structured H$_2$ photo-absorption cross section.

The dissociation of CH$_4$ leading to H$^-$+CH$_3^+$ (reaction \#~30) takes place with photons at wavelengths between 46 and 101~nm (see the dotted pink cross section in Fig.~\ref{fig:cross_sections}c). This means that Lyman $\alpha$ photons cannot be responsible for this reaction. Instead, the peak dissociation rate, located at 790~km, is driven by photons penetrating to low altitudes in the structured region of the H$_2$ photo-absorption cross section. Thus reaction \#~30 is the only CH$_4$ photo-dissociation reaction considered in this study whose reaction rate profile is sensitive to the spectral resolution of the H$_2$ photo-absorption cross section used in the model calculation. Its reaction rate profile with altitude is plotted in Fig.~\ref{fig:Pd_vs_z} in pink: the dark pink curve with a peak at 790~km is obtained when running the model with the high-resolution H$_2$ photo-absorption cross section, whereas the light pink curve is obtained using the low-resolution H$_2$ cross section. Taking into account the extra absorption occurring at low altitudes in the wings of the narrowly structured H$_2$ photo-absorption cross section is necessary to correctly determine the production peak of reaction \#~30. The other CH$_4$ dissociation profiles plotted in Fig.~\ref{fig:Pd_vs_z} are determined with the high-resolution solar spectrum, however these profiles are not sensitive to the resolution of the solar flux model used among those considered.

\subsection{Effect of the CH$_4$ inflow}
As discussed in Sect.~\ref{sec:neutral_atm}, during the final plunge INMS measured an influx of CH$_4$ \citep{Yelle2018} resulting in a constant volume mixing ratio of approximately $1.3 \times 10^{-4}$ at least down to the 1~nbar pressure level (see Fig.~\ref{fig:mixing_ratio_vs_p_INMS}). Without a high altitude source of methane, our diffusive model shows that the CH$_4$ mixing ratio only reaches values greater than $1.3 \times 10^{-4}$ at pressures higher than $10^{-7}$~bar. A large influx of methane from outside of the atmosphere has significant consequences for ion and neutral photochemistry.

We have compared the effect of the methane influx on CH$_4$ photo-ionisation and photo-dissociation rate profiles. To carry out this comparison, we run the energy deposition model using the CH$_4$ number densities reconstructed only considering diffusion (solid red curve in Fig.~\ref{fig:n_vs_p_INMS}) to compare with the production rates determined with a neutral atmosphere that includes a methane influx (dashed red curve in Fig.~\ref{fig:n_vs_p_INMS}).

The solid black curves in Figs.~\ref{fig:Pi_vs_z}(f) and \ref{fig:Pd_vs_z} show methane photo-ionisation and photo-dissociation rates, respectively, for an atmosphere without an influx of methane. The coloured curves in these plots are the rate profiles for the case with a methane influx. These figures show that the methane influx affects only the region above the peak, where the reaction rates are strongly enhanced due to the presence of additional neutral methane. The magnitude of the peak production rates are not changed by the influx. The total column methane photo-dissociation rate above the peak increases from $2.73\times 10^9$~cm$^{-2}$~s$^{-1}$ to $2.84\times 10^9$~cm$^{-2}$~s$^{-1}$ when the methane influx is considered, of which 41~\% forms CH$_3$, 53~\% forms CH$_2$, and 6~\% forms CH.

\subsection{Effect of solar spectrum and cross section resolution}
\begin{figure*}
\centering
\includegraphics[width=0.8\textwidth]{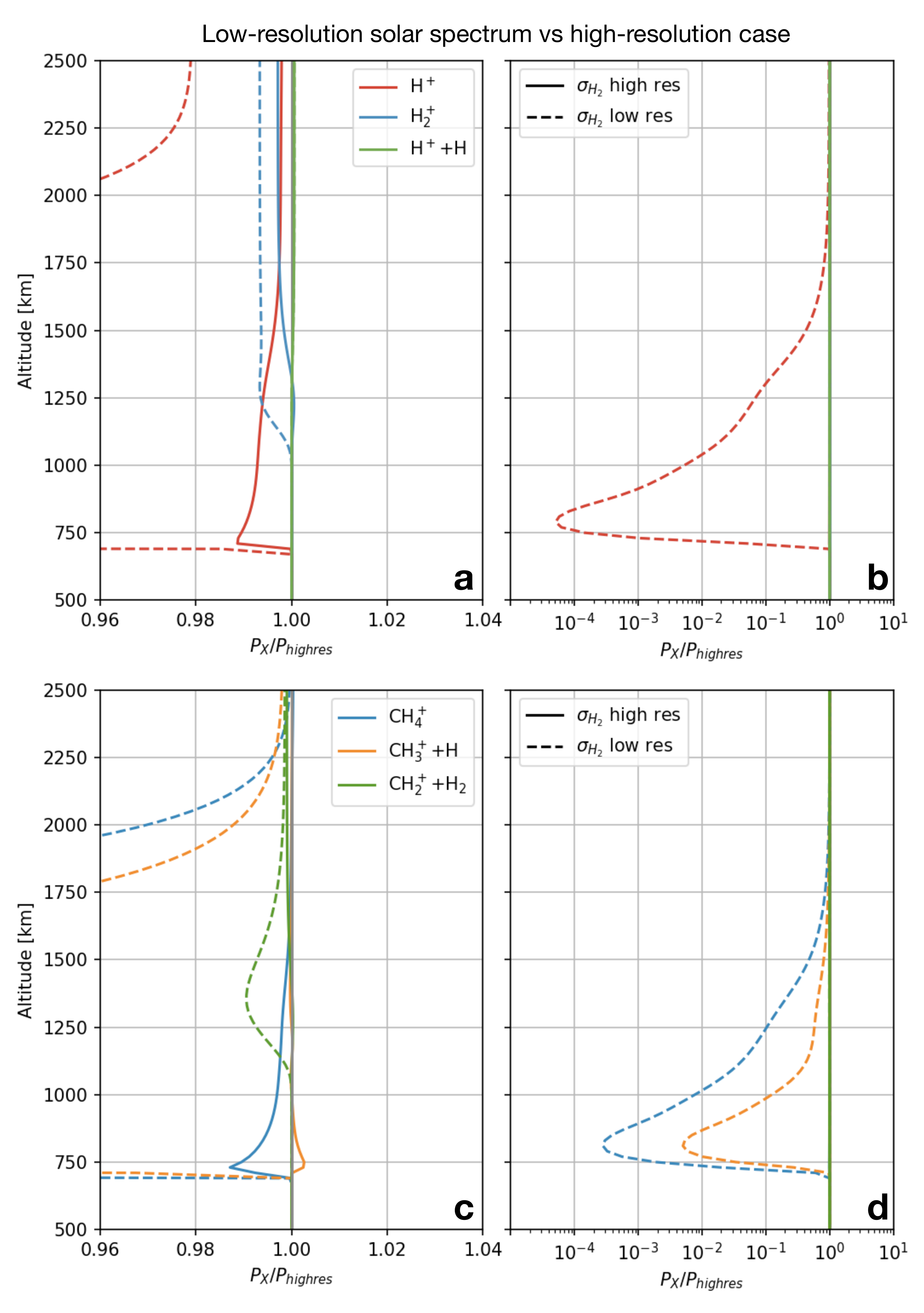}
\caption{Ratio of photo-ionisation production rates at 12 LT, showing a comparison of low-resolution cases with the high-resolution case. $P_{highres}$ is the rate calculated using the high res.~\#~1 solar spectrum (see Table~\ref{tab:construced_spectra}) and H$_2$ photo-absorption cross section. $P_X$ is the rate calculated with the low resolution solar spectrum and low resolution H$_2$ cross section (in dashed lines) or the low resolution spectrum and high resolution cross section (in solid lines). Panels a and b show photo-ionisation rates for different reactions (see Table~\ref{tab:reactionsiono}): \#4 (in red), \#1 (in blue), and \#2 (in green), whereas panels c and d show the rates for reactions \#6 (in blue), \#7 (in orange), and \#8 (in green). Panels b and d, are semi-log versions of panels a and c, respectively.}
\label{fig:ratio_Pi_lowres}
\end{figure*}

\begin{figure*}
\centering
\includegraphics[width=0.8\textwidth]{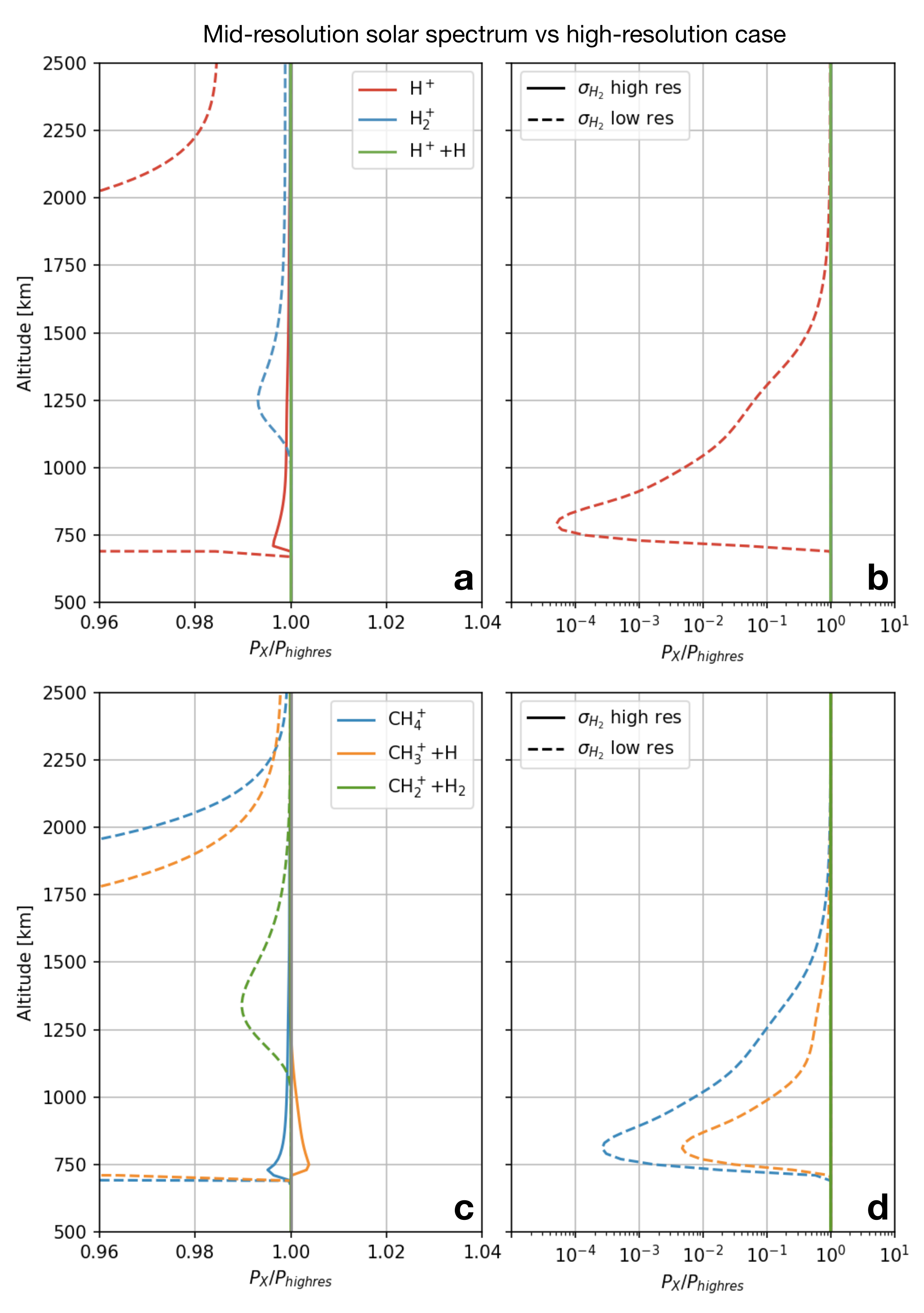}
\caption{Ratio of photo-ionisation production rates at 12 LT, showing a comparison of mid-resolution cases with the high-resolution case. $P_{highres}$ is the rate calculated using the high res.~\#~2 solar spectrum (see Table~\ref{tab:construced_spectra}) and H$_2$ photo-absorption cross section. $P_X$ is the rate calculated with the mid resolution solar spectrum and low resolution H$_2$ cross section (in dashed lines) or the mid resolution spectrum and high resolution cross section (in solid lines). Panels a and b show photo-ionisation rates for different reactions (see Table~\ref{tab:reactionsiono}): \#4 (in red), \#1 (in blue), and \#2 (in green), whereas panels c and d show the rates for reactions \#6 (in blue), \#7 (in orange), and \#8 (in green). Panels b and d, are semi-log versions of panels a and c, respectively.}
\label{fig:ratio_Pi_midres}
\end{figure*}

The reaction rates plotted in Figure~\ref{fig:Pi_vs_z} are calculated using the high resolution solar spectrum (\#~1, described in Sect.~\ref{sec:solar_flux}), along with the high resolution H$_2$ cross section (see Sect.~\ref{sec:sigma_H2}). In Figs.~\ref{fig:ratio_Pi_lowres} and \ref{fig:ratio_Pi_midres} we show the effect of running the energy deposition model using different spectral resolutions on photo-ionisation production rates at 12 LT. We note that electron impact ionisation is not affected by the use of high-resolution H$_2$ cross section and solar flux above 80~nm, as the solar photons associated with this spectral range generate photo-electrons that are too low in energy to ionise.

The curves in Figs.~\ref{fig:ratio_Pi_lowres} and \ref{fig:ratio_Pi_midres} represent the ratio of photo-ionisation production rates between model calculations using low ($\Delta\lambda=1$~nm, Fig.~\ref{fig:ratio_Pi_lowres}) and mid ($\Delta\lambda=0.1$~nm, Fig.~\ref{fig:ratio_Pi_midres}) resolution solar spectra with calculations using the high resolution solar spectrum (high res.~\#~1 in Fig.~\ref{fig:ratio_Pi_lowres}, and high res.~\#~2 in Fig.~\ref{fig:ratio_Pi_midres}, see Table~\ref{tab:construced_spectra}) and the high resolution H$_2$ cross section. Each coloured curve represents a given reaction and the line styles represent the comparison of different combinations of resolutions with the high resolution case (see Table~\ref{tab:construced_spectra}): in Fig.~\ref{fig:ratio_Pi_lowres}, the dashed lines represent the case where the low resolution solar spectrum is combined with the low resolution H$_2$ cross section, whereas the solid lines result from calculations using the low resolution solar spectrum with the high resolution H$_2$ cross section. In a similar fashion, in Fig.~\ref{fig:ratio_Pi_midres}, calculations using the mid resolution spectrum with the low resolution H$_2$ cross section are in dashed lines, and results combining the mid resolution spectrum with the high resolution cross section are shown in solid lines.

The largest deviation from the high resolution reference case is found when using the low resolution H$_2$ photo-absorption cross section, regardless of the resolution of the solar spectrum (see the dashed lines in Figs.~\ref{fig:ratio_Pi_lowres} and \ref{fig:ratio_Pi_midres}). Indeed, when using a low resolution model, there is no peak in H$^+$ production (from H) at 800~km altitude; this is reflected in panels b of Figs.~\ref{fig:ratio_Pi_lowres} and \ref{fig:ratio_Pi_midres} which show that the production from the low resolution model is equal to $6\times 10^{-5}$ times that from the high resolution model at this altitude (dashed red line). Likewise, the production of low altitude ionised hydrocarbons is not present in the models that use the low resolution cross section, as can be seen in panels d of Figs.~\ref{fig:ratio_Pi_lowres} and \ref{fig:ratio_Pi_midres}: the low resolution model values are equal to $3\times 10^{-4}$ times those from the high resolution model for CH$_4^+$ (minimum of the dashed blue line) and $6\times 10^{-2}$ times for CH$_3^+$ (minimum of the dashed orange line).

The low altitude peaks in the production of H$^+$, CH$_4^+$, and CH$_3^+$ that appear when using a high resolution H$_2$ photo-absorption cross section were also noted by \citet{Kim2014}. We find that with the additional inclusion of the large neutral methane influx recorded by INMS during the Cassini Grand Finale, the enhanced CH$_4^+$, and CH$_3^+$ production resulting from the highly structured region of the H$_2$ cross section has a significant impact up to about 1500~km altitude (see Fig.~\ref{fig:ratio_Pi_lowres}d).

The cases plotted in solid lines in Figs.~\ref{fig:ratio_Pi_lowres} and \ref{fig:ratio_Pi_midres} make use of the low or mid resolution solar spectra combined with the high resolution H$_2$ photo-absorption cross section. Interestingly, both of these cases capture the features of the fully high resolution model (i.e., high-resolution spectrum and cross section) very well for all the photo-ionisation reactions considered. Indeed, production rates determined using the low resolution solar spectrum combined with the high resolution H$_2$ cross section vary between a factor of 0.9875 and 1.0025 times those determined with the fully high resolution model. For the mid resolution solar spectrum combined with the high resolution cross section, the calculated production rates vary between 0.995 to 1.004 times those from the fully high resolution model. The largest differences occur at and just below the altitude of the production peaks. Since we are comparing between the same solar spectra at different spectral resolutions in each case -- low res.\,spectrum vs high res.~1 (see Table~\ref{tab:construced_spectra}) in Fig.~\ref{fig:ratio_Pi_lowres}, and mid res.\,spectrum vs high res.~2 in Fig.~\ref{fig:ratio_Pi_midres} -- these differences are only due to resolution of the solar spectrum. Hence, in the cases modelled in this paper, it transpires that incorporating a high resolution H$_2$ photo-absorption cross section into energy deposition models is far more important than making use of a high resolution solar spectrum.

\section{Conclusions} \label{sec:conc}
The aim of this study has been to determine the ionisation rates and CH$_4$ photo-dissociation rates in Saturn's equatorial upper atmosphere at the time and location of the Cassini final plunge. For this purpose, we have reconstructed upper atmosphere profiles of neutral temperature and major neutral densities, based upon measurements taken by INMS during Cassini's final plunge through the atmosphere. These in-situ INMS measurements were combined with previously measured CIRS limb scans and UVIS stellar occulations using a diffusion model. The resulting neutral composition contains large quantities of CH$_4$ at high altitudes, as directly measured by INMS, which are consistent with an inflow that could originate from the rings \citep{Yelle2018}. We show that the inflow does not affect the magnitude of the peak methane photo-ionisation or photo-dissociation rates, but causes a large enhancement of the rates above the peak. By fitting the results of our diffusion model to the final plunge measurements, we derive a He volume mixing ratio at the 1 bar pressure level of between 0.120 and 0.134, depending on the temperature profile used.

In addition to the details of the neutral atmosphere, we also constructed solar spectra composed to match the conditions found during the the final plunge, i.e. quiet solar conditions during solar minimum. Using values of solar F10.7 and Lyman $\alpha$ flux recorded on the day of the final plunge, we have obtained a number of solar observations taken under similar conditions. These solar measurements have allowed us to construct solar spectra at three different spectral resolutions in order to test the effect of resolution on the energy deposition model results. Indeed, since our model contains H$_2$ photo-absorption cross sections with spectral resolution up to $1\times 10^{-3}$~nm, we would ideally use a solar spectrum of equivalent resolution. The ``high-resolution'' solar spectrum used in this study is from the SUMER instrument on board SOHO and has $\Delta \lambda=4\times 10^{-3}$~nm between 67 and 152~nm.

As previous studies at Saturn \citep{Kim2014} and Jupiter \citep{Kim1994} have shown, using a high-resolution H$_2$ photo-absorption cross section that includes the very narrow absorption lines of the Lyman, Werner, and Rydberg bands results in the prediction of an additional layer of hydrocarbon ions at the bottom of the ionosphere. Indeed, we find a layer of hydrocarbon ion production near 800~km altitude (see Sect.~\ref{sec:ionisation_rates}) when using the high resolution H$_2$ cross section in the energy deposition model. Our calculated photo-ionisation rate profiles differ by less than $\pm 1.25$\% when using the low-resolution ($\Delta \lambda=1$~nm) or mid-resolution ($\Delta \lambda=0.1$~nm) solar spectra, compared to the calculations carried out with the high-resolution spectra. This indicates that as long as the full resolution H$_2$ photo-absorption cross section is used, it is much less important to also include a high resolution solar spectrum in energy deposition models of the upper atmosphere of Saturn. To this statement, we should apply the caveat that we do not have access to a solar spectrum with as high a spectral resolution as the H$_2$ photo-absorption cross section that we use. It is possible that our high-resolution spectrum is still too coarse to fully capture all of the effects on energy deposition of the highly-structured region of the H$_2$ cross section.

Since the INMS measurements during Cassini's final plunge have shown higher levels of methane than have previously been measured, we have included in this study a calculation of CH$_4$ dissociation rates. We thus find production rates of methane fragments that peak near 750~km, with large quantities of methane fragments produced throughout the thermosphere, with consequences to models of neutral photochemistry.

\section*{Acknowledgements}
Work at Imperial College London was supported by the UK Science \& Technology Facilities Council (STFC) under grants ST/N000692/1 and ST/N000838/1. We would like to thank Werner Curdt very warmly for having provided us with the high-resolution SOHO/SUMER solar spectrum used in this study. We are grateful to the TIMED/SEE team, and the Whole Heliosphere Interval (WHI) team for providing us with their solar flux data sets. We would like to thank Luke Moore very much for providing us with the INMS H$_2^+$ dataset.

\appendix

\section{Diffusion model} \label{sec:appendix}
This appendix describes the diffusion model used to reconstruct the neutral density profiles at the time of the final plunge. It is based upon \citet{Koskinen2018}.

Beginning at the 1~bar pressure level (at which the altitude is set to $z=0$ by convention), the following integral allows us to calculate the partial pressure $p_s$ of each neutral species $s$ at altitude $z$
\begin{equation}\label{eqn:partial_pressure}
p_s(z) = p_{s0}\exp\left[-\int_0^z \! \frac{\mathrm{d}z'}{1+\Lambda}\left(\frac{1}{H_s}+\frac{\Lambda}{H}\right)\right],
\end{equation}
where $p_{s0}$ is the partial pressure of species $s$ at $z=0$~km altitude, $H_s=kT/m_sg$ is the scale height of species $s$, $H=kT/mg$ is the scale height of the atmosphere, where $m$ is the mean molecular weight. $\Lambda$ is the mixing parameter
\begin{equation}
\Lambda = \frac{K_{zz}(z)}{D_s(z)},
\end{equation}
where $K_{zz}$ is the eddy diffusion coefficient, and $D_s$ is the molecular diffusion coefficient of species~$s$.

The eddy diffusion coefficient is parameterised as a function of pressure as follows \citep{Koskinen2018}:
\begin{equation}
K_{zz}(p) = \frac{K_0 K_{\infty} (p_0/p)^{\gamma}}{K_0(p_0/p)^{\gamma} + K_{\infty}},
\end{equation}
where $p_0 = p(z=0) = 1$~bar, $K_0 = 2.5$~m$^2$~s$^{-1}$ \citep{Vervack2015}, and $K_{\infty} = 1.01\times 10^8$~m$^2$~s$^{-1}$ and $\gamma = 0.32$ \citep{Koskinen2018}. $K_{\infty}$ and $\gamma$ are constrained by the CH$_4$ profile from the UVIS occultation (ST14M10D03S7). The molecular diffusion coefficient of species $s$ is given by:
\begin{equation}
\frac{1}{D_s} = \sum_{t\neq s} \frac{x_t}{D_{st}},
\end{equation}
where $x_t$ is the volume mixing ratio of species $t$, and $D_{st}$ is the binary diffusion coefficient of species $s$ through species $t$. We use the following parameterisation of the binary diffusion coefficient:
\begin{equation}
D_{st}(p,T) = A_{st}f_{st}\frac{T^{\alpha}}{p},
\end{equation}
where $A_{st}$ and $\alpha$ are empirical coefficients taken from \citet{Marrero1972}, except for the case of the diffusion of H--CH$_4$, where Equation (15.29) from \citet{Banks1973} is used. The function $f_{st}$ is dependent on temperature. Values for $A_{st}$, $\alpha$, and $f_{st}$ are given in Table~\ref{tab:molecular_diffusion}.

\begin{table}
\centering
\begin{threeparttable}
\caption{Molecular diffusion coefficients}
\begin{tabular}{lccc}
\toprule
Species      & $A_{st}$                            & $\alpha$ & $f_{st}$\\
             & [bar~cm$^2$~s$^{-1}$~K$^{-\alpha}$] &          & \\
\midrule
H$_2$--H      & $1.14\times 10^{-4}$                & 1.728    & 1 \\
H$_2$--He     & $2.74\times 10^{-2}$                & 1.510    & $\left(15.5 - \ln(T)\right)^{-2}$ \\
H$_2$--CH$_4$ & $3.17\times 10^{-5}$                & 1.765    & 1 \\
H--He         & $1.44\times 10^{-4}$                & 1.732    & 1 \\
H--CH$_4$     & $2.15\times 10^{-5}$                & 1.5      & $\left(x_{\text{H}}+x_{\text{CH}_4}\right)^{-1}$ \\
He--CH$_4$    & $3.17\times 10^{-5}$                & 1.750    & 1 \\
\bottomrule
\end{tabular}
\begin{tablenotes}
\small
\item Notes: Empirical values from \citet{Marrero1972}, apart from H--CH$_4$ where a simple expression from \citet{Banks1973} is used (Equation 15.29).
\end{tablenotes}
\label{tab:molecular_diffusion}
\end{threeparttable}
\end{table}

We construct a grid of 2770 altitude points with a spacing of 0.01 scale heights over which to integrate Equation~\ref{eqn:partial_pressure}. The gravitational field is determined using the expression of the potential $\Phi$ in Equation~\ref{eqn:potential}: $\overrightarrow{g} = \nabla \Phi$. The temperature profiles used are those we previously labelled composites A and B (shown in Fig.~\ref{fig:Tsource}), converted from a potential grid to the altitude grid using Equation~\ref{eqn:potential} at a planetographic latitude of 9$^{\circ}$N. This latitude was chosen since it corresponds to the lowest reaches of the INMS final plunge measurement.

The algorithm to obtain profiles of the neutral species is as follows. At a given altitude grid point (beginning at the lowest altitudes), we integrate Equation~\ref{eqn:partial_pressure} to obtain the partial pressures of each of the neutrals. The total pressure at this altitude point can then be determined by summing the partial pressures: $p(z) = \sum_s p_s(z)$, and hence we calculate the volume mixing ratio of each species $s$: $x_s(z) = p_s(z)/p(z)$, which allows us to obtain the mean molecular weight $m(z)$. In such a way, we obtain volume mixing ratios as a function of altitude and pressure for each of the four neutral species. 

\bibliographystyle{elsarticle-harv} 
\bibliography{references}

\end{document}